\documentclass[fleqn,usenatbib]{mnras}

\usepackage{newtxtext,newtxmath}

\usepackage[T1]{fontenc}

\DeclareRobustCommand{\VAN}[3]{#2}
\let\VANthebibliography\thebibliography
\def\thebibliography{\DeclareRobustCommand{\VAN}[3]{##3}\VANthebibliography}

\usepackage{graphicx}	
\usepackage{amsmath}	
\usepackage{booktabs} 

\usepackage[T1]{fontenc}
\usepackage{xcolor}

\def\magenta\magenta{\color{magenta}}

\newbox\grsign \setbox\grsign=\hbox{$>$} \newdimen\grdimen \grdimen=\ht\grsign
\newbox\simlessbox \newbox\simgreatbox
\setbox\simgreatbox=\hbox{\raise.5ex\hbox{$>$}\llap
     {\lower.5ex\hbox{$\sim$}}}\ht1=\grdimen\dp1=0pt
\setbox\simlessbox=\hbox{\raise.5ex\hbox{$<$}\llap
     {\lower.5ex\hbox{$\sim$}}}\ht2=\grdimen\dp2=0pt
\def\simgreater{\mathrel{\copy\simgreatbox}}

\newbox\simppropto
\setbox\simppropto=\hbox{\raise.5ex\hbox{$\sim$}\llap
     {\lower.5ex\hbox{$\propto$}}}\ht2=\grdimen\dp2=0pt

\title[Insight into the birth of $\omega$ Cen's MPs]{Chemical tagging with APOGEE, MUSE, and HST: constraints on the formation of $\omega$ Centauri}

\author[Andrew C. Mason et al.]{\noindent
Andrew C. Mason$^{1,2}$, Ricardo P. Schiavon$^{1}$\thanks{E-mail: r.p.schiavon@ljmu.ac.uk},  Sebastian Kamann$^{1}$, Verne V. Smith$^3$, Danny Horta$^4$ \newauthor 
Borja Anguiano$^{5,6}$, Katia Cunha$^{7}$, Szabolcs~M{\'e}sz{\'a}ros$^{8,9}$, Steven R. Majewski$^6$, Robert W. O'Connell$^{6}$, 
\newauthor 
Carlos Allende Prieto$^{10}$, Sara Saracino$^{1,11}$\\
$^{1}$Astrophysics Research Institute, Liverpool John Moores University, 146 Brownlow Hill, Liverpool, Merseyside, L3 5RF\\
$^{2}$Institute of Systems, Molecular, and Integrative Biology, University of Liverpool, Biosciences Building, Crown Street, Liverpool, Merseyside, L69 7BE\\
$^{3}$NSF's NOIRLab, Tucson, AZ 85719, USA\\
$^{4}$Institute for Astronomy, University of Edinburgh, Royal Observatory,
Blackford Hill, Edinburgh EH9 3HJ, UK\\
$^{5}$Centro de Estudios de F\'isica del Cosmos de Arag\'on (CEFCA), Plaza San Juan 1, 44001, Teruel, Spain\\
$^{6}$Department of Astronomy, University of Virginia, Charlottesville,
VA, 22904, USA\\
$^{7}$Steward Observatory, University of Arizona, 933 North Cherry Avenue, Tucson, AZ 85721-0065, USA\\
$^{8}${ELTE E\"otv\"os Lor\'and University, Gothard Astrophysical Observatory, 9700 Szombathely, Szent Imre H. st. 112, Hungary}\\
$^{9}${MTA-ELTE Lend{\"u}let "Momentum" Milky Way Research Group, Hungary}\\
$^{10}$Instituto de Astrofisica de Canarias, Via Lactea s/n La Laguna 38205 La Laguna Spain\\
$^{11}$INAF – Osservatorio Astrofisico di Arcetri, Largo E. Fermi 5, 50125 Firenze, Italy}

\date{Accepted XXX. Received YYY; in original form ZZZ}

\pubyear{2015}

\begin{document}
\label{firstpage}
\pagerange{\pageref{firstpage}--\pageref{lastpage}}
\maketitle

\begin{abstract}
A plethora of evidence suggests that $\omega$ Centauri ($\omega$ Cen) is the nuclear star cluster of a galaxy that merged with the Milky Way in early times. 
We use APOGEE, Gaia, MUSE, and HST data supplemented by galaxy chemical evolution models to place constraints on the assembly and chemical enrichment history of $\omega$ Cen. 
The APOGEE data reveal three stellar populations occupying separate loci on canonical chemical planes. 
One population resembles metal-poor halo field stars (P1), a second shows light-element abundance anti-correlations typical of metal-poor globular clusters (IM), and a third population (P2) is characterised by an extreme "second-generation" abundance pattern.
Both P1 and P2 populations cover a broad range of metallicity, consistent with extended histories of bursty star formation (SF), which is also evident from their light- and $\upalpha$-element abundance patterns. 
Conversely, the IM stars exhibit a narrow metallicity spread, combined with the  Al-Mg, Na-O, and C-N anti-correlations common to metal-poor Galactic globular clusters. Moreover, these three populations alone seem to account for the distribution of $\omega$~Cen stars in the chromosome map.
We discuss these findings in  context of a scenario according to which $\omega$~Cen formed by a combination of {\it in situ} SF within the host galaxy (P1), followed by the spiralling in of gas-rich globular clusters (IM), leading to another burst of SF (P2). 
We perform a robust comparison of the chemical composition of $\omega$~Cen with those of  halo substructures well represented in APOGEE DR17, finding no chemical associations to a high confidence level.
\end{abstract}




\section{Introduction}
\label{sec:intro}

In the prevailing $\Lambda$CDM cosmogony, the assembly of the Galaxy was partly due to its accretion of many so-called `building blocks' (i.e., dwarf galaxies) during earlier cosmic epochs. In this vein, advancements in the field of Galactic archaeology have led to a number of associations between Galactic globular clusters (GCs) and the debris that comprise the Milky Way's (MW) stellar halo \citep[e.g.,][]{massari_gc,horta2020}. 
Such associations find support in the theoretical expectation that building blocks accreted at early cosmic epochs are survived by at least some members of their GC systems at $z=0$ \citep[e.g.,][]{zinn_searle_78, kruijssen_19}. 
Furthermore, nuclear star clusters (NSCs), which are speculated to form by a combination of {\it in-situ} star formation and the spiralling in of GCs through dynamical friction \citep[see][and references therein]{neumayer_nsc_review}, are observed at the centre of potential of most galaxies and can also survive their hosts after major mergers. Perhaps the best known example is NGC~6715 (M54) which is suggested to be the NSC of the Sagittarius dSph, a dwarf galaxy currently in the process of merging with the Milky Way \citep{ibata_m54}.

$\omega$ Centauri ($\omega$ Cen; NGC~5139) is the most massive ($M=3.55\times10^{6}~{\rm{M_\odot}}$; \citealp{baumgardt_gcs}) of the Milky Way's GCs. 
It has been shown to host multiple stellar populations \citep[MPs; see][and references therein]{renzini2015,bastian_gc_review},  which are obvious both {from photometric} \citep[e.g.,][]{milone_atlas_chm,nitschai_chm} and {spectroscopic} \citep[e.g.,][]{jp10, marino_sodium, alvarez_garay_wcen} { evidence}. 
$\omega$ Cen's stars exhibit a broad spread in metallicity \citep[e.g.,][]{pancino2000,frinchaboy_mdf, haberle_23} with its metallicity distribution function (MDF) showing multiple distinct peaks in [M/H] \cite[e.g.,][]{villanova_amr,jp10,alvarez_garay_wcen}. This indicates that it formed its MPs over an extended star formation history (SFH) that was likely characterised by multiple {bursts}. 
{These properties set $\omega$~Cen far apart from} ``normal'' Galactic GCs, which are notionally characterised as a mono-metallic stellar populations.


$\omega$ Cen also has a retrograde, coplanar orbit relative to the Milky Way disk \citep[e.g.,][]{dinescu99_velocities, majewski_wcen_retrograde}, which has naturally led to conjecture as to its origin in the context of the hierarchical assembly of the Milky Way, and not just the physics responsible for the formation of its MPs. 
Earlier work speculated that $\omega$ Cen may be the nucleated remnant of a galaxy which originally resembled the massive dSphs we see in the Local Group today. According to this scenario, such a galaxy would have then been captured by the Milky Way \citep[e.g.,][]{bekki_wcen} and gradually stripped over many passages, in a process similar to that currently undergone by the Sagittarius dSph. 

Such {antecedents might} explain its retrograde orbit and structural parameters, which place it on the border between the loci occupied by GCs and the ultra-compact dwarfs (UCDs) in the luminosity-size relation, \citep[e.g.,][and references therein]{tolstoy-09}. Indeed, it is thought that UCDs form by the same process, though their progenitors are thought to be more massive {than $\omega$~Cen} \citep{pfeffer13, pfeffer14}.

While an enormous amount of work has gone into obtaining data from $\omega$ Cen and constraining its properties, it was not until the advent of {\it Gaia} \citep{gaia16} {and massive spectroscopic surveys} that these data could be placed in context of a representative sample of stars within the Galaxy. 
Combination of detailed chemical compositions and radial velocities of individual stars from surveys such as LAMOST, APOGEE, and GALAH \citep[][]{lamost_12,apo17,galah} with astrometric information from {\it Gaia} has enabled the construction of a rich multi-dimensional chemo-kinematic dataset. 


The above data prompted speculation as to the association of $\omega$ Cen with recently identified halo substructures. \cite{massari_gc}, \cite{forbes2020_mw_assembly}, and \cite{pfeffer2021_nscs} proposed that $\omega$ Cen was the nuclear star cluster of the Sausage/Gaia Enceladus' progenitor galaxy \citep[S/GE;][]{belokurov_sausage,haywood2018,hayes18_gse,helmi_ges,mackereth2019}, which underwent a major merger with the Milky Way some $\simeq10~{\rm Gyr}$ ago.   
Alternatively, \citealp{Myeong_sequoia} advocate that $\omega$ Cen is instead associated with the `Sequoia' remnant, {which in turn has been claimed to be} associated with the bulge GC FSR 1758 by \cite{barba19_sequoia}. 

It is thought that there is a specific mass range in the scaling relation between $M_{\rm host}$ and $M_{\rm NSC}$ within which NSCs formed by an almost 50/50 mixture of {\it in-situ} star formation and the inspiralling of a number of its host's globular clusters \citep{neumayer_nsc_review, fahrion_nscs_19}. If $\omega$ Cen does originate from the Sausage/Gaia-Enceladus, estimates for the mass of this system \citep[e.g.,][]{mackereth2020,limberg_sausage} and the measured mass of $\omega$ Centauri place it firmly within the mass range considered for systems thought to have formed by this mechanism.

In this paper, we present an analysis of the properties of $\omega$ Cen using a combination of stellar abundances, astrometry, and HST photometry. In $\S$\ref{sec:data}, we describe the data used for our study, and our procedure for selecting individual stellar populations within the sample, as well as the procedure we used to construct the so-called `chromosome map' of $\omega$ Cen.
In $\S$\ref{sec:chemistry}, we show the detailed chemistry of the multiple populations in $\omega$ Cen and provide a quantitative description of their properties, and speculate on the origin of their abundance patterns.
By {matching our HST and APOGEE catalogs}, we tie our analysis into the wider observational state of play by directly linking the loci stars occupy on the ``chromosome map'' to their detailed stellar abundance patterns.
In $\S$\ref{sec:gce_modelling} we fit galaxy chemical evolution models to two of the three populations we identify in $\omega$ Cen. 
Finally, in $\S$\ref{sec:discussion}, we provide a  {speculative} formation scenario for the cluster, and perform a detailed statistical comparison between what we identified as the `field' population of $\omega$ Cen's progenitor and purported substructure in the Galaxy's stellar halo.
\section{Data and Methods}\label{sec:data}

Our analysis is based on an amalgamation of data from three different sources.  Detailed chemical compositions for 1,756 $\omega$ Cen stars are extracted from the \cite{vac_paper} Value Added Catalogue (henceforth, simply VAC) of Galactic GC stars from the 17$^{\rm th}$ data release (DR17) by the Apache Point Observatory Galactic Evolution Experiment \citep[APOGEE][]{apo17,apo22}. This catalogue is supplemented by {\it Gaia} astrometry \citep{gaia_dr3}, providing coordinates and proper motions alongside radial velocities from APOGEE.

Additional spectroscopic and multi-band photometric data come from the oMEGACat catalogue \citep{haberle_23,haberle_24}, which combines spectroscopy from the ESO/VLT Multi Unit Spectroscopic Explorer \citep[MUSE,][]{muse_1,muse_2} with PSF photometry from the Hubble Space Telescope. The latter data are based on Advanced Camera for Surveys Wide Field Channel (ACS/WFC) and Wide Field Camera 3 UVIS Channel (WFC3/UVIS) covering the half-light radius of $\omega$ Cen ($R\approx5'$).

In this section, we describe the data, our crossmatch between the VAC and oMEGACat, the quality cuts we perform on both samples, our construction of the so-called ``chromosome map'' (ChM) to reproduce that by \cite{nitschai_chm}, and finally our method to split the VAC sample into different stellar populations on the basis of their chemical {compositions}.

\subsection{APOGEE data for $\omega$~Cen members}\label{subsec_vac}

This paper combines the latest data release (DR17; \citealp{apo17, apo22}) of the SDSS-III/IV \citep{eisenstein_sdss_11, blanton_sdss_iv_15} and APOGEE survey \citep{apo17, apo22} with distances and astrometry derived from the third data release of the {\it Gaia} survey \citep{gaia_dr3}.  
The APOGEE DR17 catalogue {adopted ({\tt allStar-dr17-synspec\_rev1.fits}) comprises  stellar parameters and high precision elemental abundances for up to} 20 species as well as radial velocities for approximately $\sim 700,000$ stars in total, within the Milky Way and a number of its satellites and GCs.

Elemental abundances and radial velocities were obtained from the analysis of high-resolution near-infrared spectra of hundreds of thousands of stars in both hemispheres, observed with the Apache Point Observatory 2.5m Sloan telescope \citep{gunn_06} and the Las Campanas Observatory 2.5m Du Pont telescope \citep{bowen_dupont}.  
These spectra were obtained using twin high efficiency multi-fiber NIR spectrographs assembled at the University of Virginia, USA \citep{wilson_19_apogee}. A technical summary of the overall SDSS-IV experiment can be found in \cite{blanton_sdss_iv_15}.

Further in-depth information on the APOGEE survey, data, and data reduction pipeline can be found in \cite{apo17}, \cite{jonsson_20} \& \cite{holtzman_18}, and \cite{nidever_apogee_pipe}, respectively. The APOGEE Stellar Parameters and Abundances Pipeline (ASPCAP) is described in \cite{aspcap_16}.

Chemical composition data based on earlier APOGEE data releases were presented for a number of Galactic GCs \citep{meszaros2015,schiavon2017b, meszaros2019, meszaros2020, meszaros2021}.  However, on a GC-by-GC basis {sample sizes, spatial coverage, and magnitude limits vary substantially}.  
Furthermore, prior APOGEE data releases lack robust estimates of star-by-star GC membership probabilities. 
To address this issue, \cite{vac_paper} produced the SDSS/APOGEE Value Added Catalogue of Galactic Globular Cluster (GC) Stars (VAC). This VAC is the result of a sweeping search of the APOGEE DR17 catalogue for {likely} GC members using a set of membership criteria, leveraging precise astrometry {(positions and proper motions) from {\it Gaia}, with radial velocities and chemical compositions from} APOGEE.

In this paper, we concern ourselves primarily with a subset of the APOGEE VAC, namely giants located in $\omega$ Cen. The sample of $\omega$ Cen stars analysed within this study is defined by the following set of criteria:
\begin{enumerate}
    \item \texttt{GC\_NAME}=NGC5139
    \item $p_{\rm{\omega Cen}}>0.5$
    \item $\log g<3.6$
    \item 3500 K < $T_{\rm{eff}}$ < 4500K
    \item S/N > $70~{\rm{pixel}}^{-1}$,
\end{enumerate}
where $p_{\mathrm{\omega Cen}}$ is the $\omega$~Cen membership probability (from the VAC) and the other parameters have their usual meaning.  These criteria yielded {1,555 unique stars} in total. When data for other GCs are described in the analysis, they are subject to identical criteria on a GC-by-GC basis.

\subsection{Complementary HST and MUSE data for \texorpdfstring{$\omega$}{omega} Cen stars}\label{subsec_hst_muse}

Complementary HST photometry for APOGEE stars is derived from the oMEGACat catalogue \citep{haberle_23, haberle_24}, comprising both MUSE and HST observations of individual stars within $\omega$ Cen, out to the half-light radius {\citep[$R\simeq4.65'$][]{baumgardt_hilker2018}}. We applied the same quality cuts (QC) as in \cite{nitschai_chm}, whereby we select red giants comprising the MUSE QC sample that are also present in the HST QC sample from \cite{haberle_24}. Thus, we are left with 10,850 stars satisfying the following conditions:
\begin{enumerate}
    \item Present in the HST QC with $m_{F625W}<17~{\rm mag}$
    \item Measurements in F625W, F435W, F275W, F336W, and F814W
\end{enumerate}

Any star in $\omega$ Cen also observed as part of the {\it Gaia} mission has its source ID present in the oMEGACat catalogue, making the {match} to the VAC {sample} trivial. There are 135 stars present in oMEGACat satisfying the above conditions, that are also included in the VAC.

\subsubsection{Constructing the \texorpdfstring{$\omega$}{omega} Cen chromosome map}\label{subsubsec_chm}

To briefly summarize, so-called ``chromosome maps'' have become a valuable diagnostic tool for identifying stellar populations with abundance anomalies in  GCs. Combining ultraviolet and optical multi-band photometry, they are very useful for the characterization of multiple populations in GCs for which spectroscopic abundances of large samples of member stars are not available which is often the case.


Constructing a ChM requires a given CMD to be verticalised at fixed colour, such that characteristic variations in colour of individual stars with respect to the run of the red giant branch may be computed star-by-star. Judicious choice of filter combinations sensitive to physical properties, such as helium or light-element abundance and effective temperature, reveals different populations that may not be obvious from a glance of a standard CMD.

The colours of choice are
$\rm{m}_{F275W}-\rm{m}_{F814W}$ and $\rm{C}_{\rm{F275W,~F336W,~F435W}}$, where the latter is defined according to:
\begin{equation}\label{pseudocolor_equation}
\begin{split}
{\rm{C}}_{{\rm F275W,~F336W,~F435W}}=\\
(m_{\rm F275W}-m_{\rm F336W})- (m_{\rm F336W}-m_{\rm F435W}).
\end{split}
\end{equation}
The reference magnitude for each CMD is $m_{\rm F814W}$. The pseudocolor $C_{\rm F275W,~F336W,~F435W}$ is adopted as it is a potent tracer of the degree of CNO-process enrichment experienced by stars. This is due to the fact that this combination of filters encompasses the OH molecular band; the NH band; and the CH and CN bands in F275W, F336W, and F435W, respectively \citep{milone2012_47tuc, milone2015_HUGGS}.

We follow the procedure from \cite{nitschai_chm}, briefly described here, but refer readers to Appendix C of that paper for detailed instructions. We start by selecting a sample of red giant stars meeting the quality cuts described in $\S$\ref{subsec_hst_muse}.
Our procedure is mostly identical to \cite{milone_chm}, except that we use different photometric filters \citep[][used F438W instead of F435W]{milone_chm}. As stated in that paper, the metallicity spread in $\omega$ Cen requires constructing several fiducial lines for sequences on the CMD with different [M/H]. 
We thus break the sample down by metallicity sub-group, by applying a 1D Gaussian Mixture Model (GMM) to the [M/H] distribution. 
We found 11 to be the number of components with the lowest Bayesian Information Criterion (BIC, after running up to 30), in agreement with \cite{nitschai_chm}.
We then combine labels from the GMM procedure to derive three samples corresponding to the metal-poor, metal-intermediate, and metal-rich stars by visual inspection.

The next step is then to derive the fiducial lines for each metallicity group in both CMDs. 
We start by describing the method applied to the CMD based on the $m_{\rm F275W}-m_{\rm F814W}$ colour. For the metal-poor population, we used LOWESS smoothing \citep{cleveland79_lowess} in order to compute the difference at fixed magnitude for individual stars between their color values and the LOWESS curve, $\delta m$. 
Then, the fiducials corresponding to the position of the $4^{\rm th}$ and $96^{\rm th}$ percentiles of the distribution of $\delta m$ were simply constructed by adding $\pm~2~\sigma$ to the LOWESS-determined median. 
For the intermediate and metal-rich populations, the red giant branches are well separated, so that single LOWESS fiducial lines are calculated for $m_{\rm F275W}-m_{\rm F814W}$, and $\delta m$ is derived from these for each star 

The procedure differed slightly {when using the pseudocolor} $\rm{C}_{\rm{F275W,~F336W,~F435W}}$.  That is because the giant branches of the metal-poor and intermediate population are well separated in $m_{\rm F275W}-m_{\rm F814W}$, but not in $\rm{C}_{\rm{F275W,~F336W,~F435W}}$.  
Therefore, $\Delta_{\rm{F275W,~F336W,~F435W}}$ values are calculated on the basis of fiducial lines, corresponding to the $4^{\rm th}$ and 96$^{\rm th}$ percentiles for {\it i)} a combination of the metal-poor and intermediate stars, and {\it ii)} the metal-rich stars. Using these fiducial lines for both CMDs, we derived star-by-star values of $\Delta_{\rm F275W,~F336W,~F435W}$ and $\Delta_{\rm F275W,~F814W}$, calculated for each colour using Eqs. C1-C11 and conditions from Table 3 in \cite{nitschai_chm}.

\subsection{Chemically tagging the multiple populations of \texorpdfstring{$\omega$}{omega} Centauri}\label{subsec_tagging}

In order to {identify} the multiple populations (MPs) of $\omega$ Cen on the basis of their chemistry, we use $k$-means clustering applied to a parameter space {consisting} of the following stellar abundances of stars in the sample: [Fe/H], [Mg/Fe], [Si/Fe], [Al/Fe], and [Mn/Fe]. 
We omit C, N, and O because their abundances are strongly affected by evolution along the RGB. 
After applying the algorithm from $k=1$ to $k=10$ clusters, we determine the best-fitting $k$ by evaluating the gap statistic for each iteration and picking the value corresponding to the number that minimises the gap statistic.

The elements Mg and Al are thought to be contributed to the star-forming gas by SN II explosions \citep{woosley95, portinari98, chieffi04, kobayashi06, nomoto13}, with Al having yields that strongly depend on metallicity \citep{weinberg_scaling}. Si is predominantly formed in SN~II, but has some contribution by SN~Ia \citep{kobayashi2020}. In GCs, all three elements are  affected by the MP phenomenon, where anomalous populations tend to exhibit Al-enhancement, Mg-depletion, and {in a few cases} Si-enhancement \citep{alvarez_garay_wcen}. Mn is an Fe-peak element, produced by both SN II and SN Ia \citep{weinberg_scaling}, widely used in combination with Mg and Al to discriminate accreted from {\it in-situ} populations \citep{hawkins15_mgmn, das20_mgmn, horta21_heracles,horta_schiavon2024}.

\begin{figure}
    \includegraphics[width=\columnwidth]{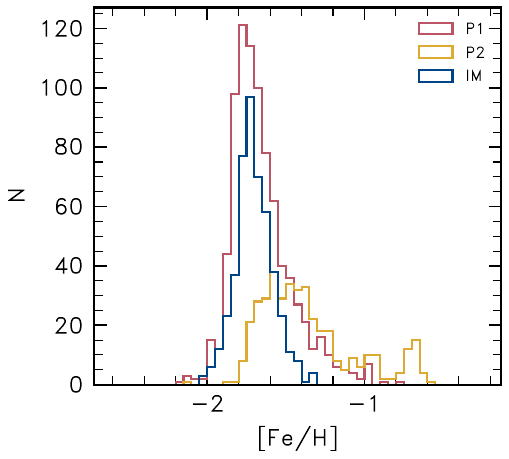}
    \caption{Metallicity distribution functions of the three populations identified by $k$-means clustering in $\omega$ Cen, which we label P1 (red), P2 (yellow) and IM (for intermediate; blue). P1 is characterised by its lower [Fe/H] and a tail towards higher [Fe/H], whereas P2 is more metal-rich and has a broader metallicity spread. Conversely, the intermediate population is characterised by a narrow dispersion in [Fe/H] at the same [Fe/H] as P1 (see Table \ref{tab:summary}).}
    \label{fig:1_mdf}
\end{figure}

Using a combination of these abundance ratios, we rescale {the numerical values} by subtracting the mean from each {stellar} abundance and then dividing the resultant number by the variance about the mean. On the basis of this rescaled dataset, we find that the data are well-described by $n=3$ clusters, {whose MDFs are shown} in Fig. \ref{fig:1_mdf}.

We label the {three} populations P1, P2, and IM. The reasons for this nomenclature {are clarified} in $\S$\ref{sec:chemistry}. {The IM population has a relatively narrow MDF, ranging from [Fe/H]$\simeq-2$ to $-1.3$, whereas P1 and P2 both present tails extending towards higher metallicity ([Fe/H]$\simgreater-1$.  The MDF of the P2 population peaks at a slightly higher [Fe/H] than P1 and IM.}


\subsection{The distribution of \texorpdfstring{$\omega$}{omega} Cen's Multiple Populations on the colour-magnitude diagram}

\begin{figure*}
    \includegraphics[width=\textwidth]{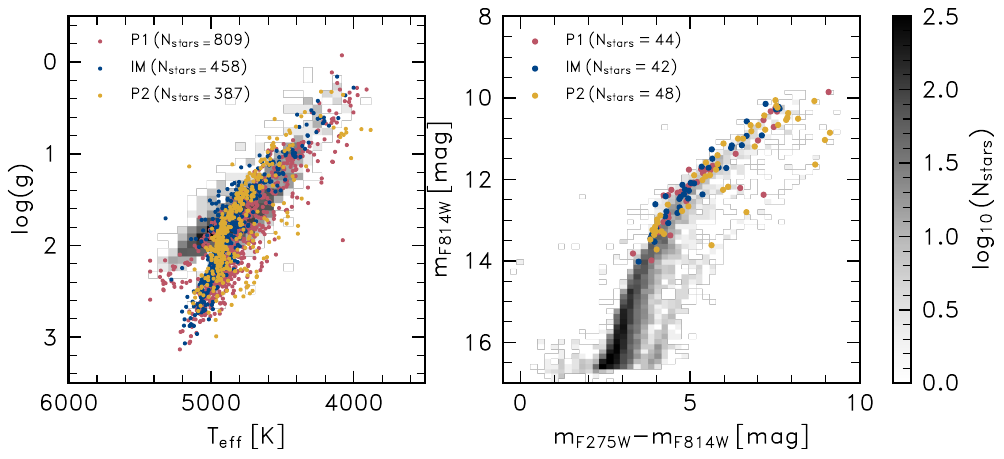}
    \caption{2d histograms, where each pixel represents the number of stars measured in {\it i)} the Kiel diagram derived from APOGEE stellar parameters adopting 50 K  and 0.1 dex bins in $T_{\rm eff}$ and $\log(g)$, respectively; and {\it ii)} the HST CMD from oMEGACat \citep{haberle_24} adopting 0.1 mag bins in both $m_{\rm F275W}-m_{\rm F814W}$ and $m_{\rm F814W}$. Coloured points, adopting the same colour scheme as Fig. \ref{fig:1_mdf}, show the chemically tagged stars belonging to each subpopulation identified in this paper (whose properties are summarised in Table \ref{tab:summary}). {Note that, while the IM RGB is relatively narrow and blue, those of the P1 and P2 populations extend towards redder colours, as expected from the MDFs in Figure~\ref{fig:1_mdf}.}}
    \label{fig:cmds}
\end{figure*}

Fig. \ref{fig:cmds} shows the distribution of chemically tagged populations in $\omega$~Cen on {\it i)} the Kiel diagram for $\omega$ Cen stars in the APOGEE VAC ({\it left}), and {\it ii)} the $m_{\rm F275W}-m_{\rm F184W}$-$m_{\rm F814W}$ CMD derived from HST photometry available in the oMEGACat catalogue ({\it right}). 

In both cases, the underlying plot is represented by a 2d histogram where each pixel represents the number of stars in each bin, and individual stars (where available, in the latter case) within each population are shown as points adopting the same colour scheme for the P1, IM, and P2 populations as seen in Fig. \ref{fig:1_mdf}.

Fig. \ref{fig:cmds} shows that P1 and P2 are, indeed, characterised by large spreads in [Fe/H] as evidenced by the broad loci they occupy on the Kiel diagram and CMD. By contrast, the IM population occupies a much narrower locus - indicative of the fact that it may have formed in a brief episode of star formation, limiting its metallicity spread.

In $\S$\ref{sec:chemistry} we place these populations on canonical chemical planes and characterise their abundance patterns.

\section{Interpreting the Abundance Patterns of \texorpdfstring{$\omega$}{Omega} Centauri's Multiple Stelar Populations}\label{sec:chemistry}

In this section, we first provide a quantitative analysis of the abundance patterns of our three populations identified by the procedure described in $\S$\ref{subsec_tagging}, and speculate on the origin of these abundance patterns { and how they shed light on $\omega$ Cen's assembly history.}

\begin{figure*}
    \centering
    \includegraphics{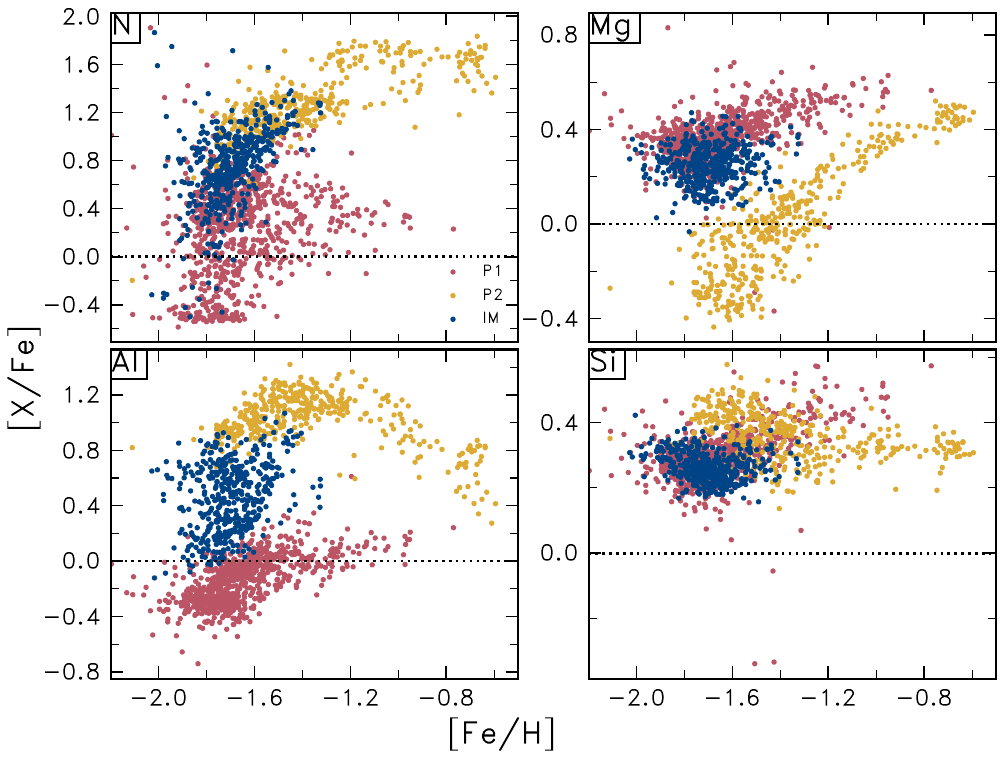}
    \caption{Distribution of $k$-means selected clusters on chemical planes using APOGEE data, where the $y$-axis values are the element abundance ratios [X/Fe] for species X in Table \ref{tab:summary} plotted as a function of [Fe/H]. Consistent with prior observations, the most obvious feature of these abundance planes is the discreteness of what we dub the P1 (red) and P2 (yellow) populations. The former is characterised by initially halo-like light-element abundances that increase over the whole range of [Fe/H], consistent with a starburst. P2 is characterised initially by heavy depletion in Mg, heavy enhancement in Al and enhancement in Si. As chemical enrichment took place in the cluster, the abundances tend back toward those characteristic of enrichment by SN~II and SN~Ia. The IM (navy) population has a narrow metallicity spread compared to P1 and P2, slight depletion in Mg (but not Si), and intermediate C, N, and Al-enhancement between P1 and P2.}
    \label{fig:6panel_abundances}
\end{figure*}

{Fig.~\ref{fig:6panel_abundances} shows the distributions of the three $\omega$ Cen populations described in Section~\ref{subsec_tagging} in various chemical planes.}
{Table~\ref{tab:summary}, lists} {\it i)} the median [X/Fe], {\it ii)} the dispersion in [X/Fe] ($\sigma_{[\rm{X/Fe}]}$), and {\it iii)} the Spearman rank correlation coefficient ($R_{S}$) computed for each abundance ratio [X/Fe] with respect to [Fe/H]. 
We also provide measurements of the median [Fe/H], and the dispersion in [Fe/H] represented by the standard deviation ($\sigma_{[{\rm{Fe/H}]}}$).

Fig.~\ref{fig:6panel_abundances} displays some striking features.  Firstly, P1 and P2 are clearly separated in these chemical planes.
Secondly, robust correlations between [X/Fe] and [Fe/H] for several elements attest to the occurrence of strong chemical evolution induced by a history of  star formation.  
Thirdly, the fact that the P1 and P2 populations are extended over {\it parallel} sequences in many of the planes indicates that these populations likely evolved in chemical detachment, suggesting that P1 and P2 stars formed in different locations, at different times, or both \citep[see discussion by][for similar considerations in the context of the $\alpha$-bimodality in galaxy discs]{mackereth2018}.
Finally, the distribution of the IM population in all chemical planes is not characterised by a significant correlation between any abundance ratios and [Fe/H].  Instead, this population has a very small metallicity spread, combined with a large spread in the abundances of some elements, such as Al, C, and N. In fact, we show in Section~\ref{sec:gce_modelling} that these abundance variations display the anti-correlations known to exist in Galactic GCs \citep{ventura_agb_yields, alvarez_garay_wcen}.    

\begin{figure*}
    \centering
    \includegraphics{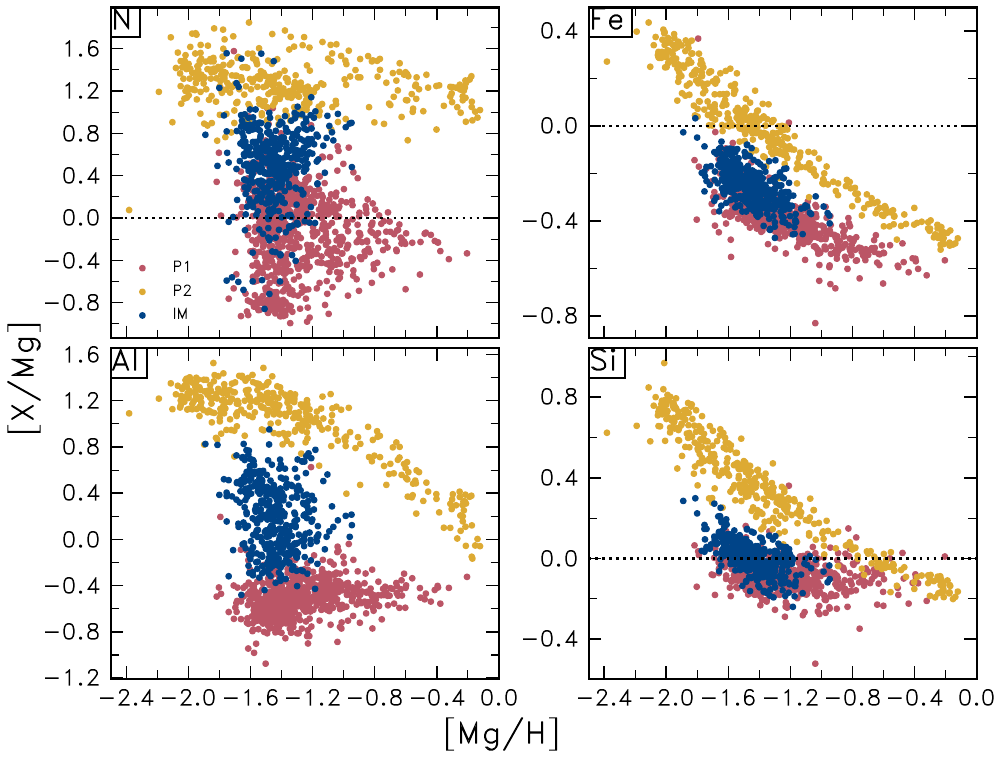}
    \caption{Distribution of $k$-means selected clusters on chemical planes, where the $y$-axis values are the element abundance ratios [X/Mg] for species X used in the clustering plotted as a function of [Mg/H]. Here, it becomes more obvious that the IM and P2 populations are far more distinct from one another. The Al-Mg plane in particular shows that there is a discontinuity in the abundance patterns around [Al/Mg]$\simeq0.8$ - it is clear that whatever produced the abundance pattern of the P2 stars must be distinct from the IM population.}
    \label{fig:6panel_abundances_mg}
\end{figure*}

\begin{table*}
    \centering
    \label{tab:summary}
    \begin{tabular}{lccccccccc}
        \toprule
        & \multicolumn{3}{c}{P1} & \multicolumn{3}{c}{IM} & \multicolumn{3}{c}{P2} \\
        \cmidrule(lr){2-4} \cmidrule(lr){5-7} \cmidrule(lr){8-10}
        X & [X/Fe] & $\sigma_{[{\rm{X/Fe}}]}$ &  $R_{{\rm{S}}}$ & [X/Fe] & $\sigma_{[{\rm{X/Fe}}]}$ &  $R_{{\rm{S}}}$ & [X/Fe] & $\sigma_{[{\rm{X/Fe}}]}$ &  $R_{{\rm{S}}}$  \\
        \midrule
        C  & 0.07 & 0.36 & 0.54 & -0.10 & 0.32 & 0.14 & -0.08 & 0.21 & 0.12 \\
        N  & 0.36 & 0.37 & 0.17 & 0.82 & 0.31 & 0.52 & 1.26 & 0.25 & 0.77\\
        Mg  & 0.36 & 0.11 & 0.58 & 0.26 & 0.08 & 0.09 & 0.00 & 0.24 & 0.84 \\
        Al  & -0.11 & 0.17 & 0.69 & 0.48 & 0.24 & 0.30 & 1.05 & 0.19 & -0.06 \\
        Si  & 0.29 & 0.09 & 0.55 & 0.26 & 0.04 & -0.13 & 0.36 & 0.07 & -0.47 \\
        Mn  & -0.35 & 0.27 & -0.37 & -0.31 & 0.26 & -0.51 & -0.38 & 0.22 & -0.18 \\
        \midrule
        Median [Fe/H] &  & -1.70 &  &  & -1.69 &  &  & -1.41 & \\
        $\sigma_{[\rm{Fe/H}]}$ &  & 0.19 &  &  & 0.12 &  &  & 0.30 & \\
        \bottomrule
    \end{tabular}
    \caption{Summary of abundance ratios and their properties for three samples.}
\end{table*}

\subsection{The abundance pattern of the P1 population}\label{pattern_p1}

The chemical compositions of the P1 population at the low metallicity end exhibit a pattern that resembles that of field stars at the same metallicity (e.g., [Mg/Fe]$\simeq0.4$ at [Fe/H]~$\simeq~-1.8$), which is consistent with the value of the high-$\upalpha$ plateau in Galactic field populations \citep[e.g.,][]{mackereth2017,horta21_heracles,horta23_halosubs} .  
Above [Fe/H]~$\approx-1.62$ there is a monotonically increasing trend of [$\upalpha$/Fe] with [Fe/H] which is seen in both Mg and Si and reflected by the high Spearman rank correlation coefficients ($R_{{\rm{S}}}$, 0.55 and 0.58, respectively) for both of those abundances as a function of [Fe/H]. 
Such a strong positive correlation between [$\upalpha$/Fe] and [Fe/H] implies that P1 {underwent an early} starburst which resulted in the chemical enrichment by CCSNe dwarfing that by SNe Ia \citep{gilmore91, weinberg_sudden_events_17, mason_knee}. 
The P1 stars also exhibit similar behaviour in [Al/Fe].  At low [Fe/H] ([Fe/H]$\simeq-2.0$), P1 shows [Al/Fe]~$\simeq-0.3$, which is consistent with what is seen in the halo field \citep[e.g.,][]{horta23_halosubs}. 
In addition, [Al/Fe] is strongly correlated with [Fe/H]  ($R_{\rm{S}}$=0.69). As in the case of $\upalpha$ elements, a significant fraction of Al is produced in CCSNe, so that this trend is further evidence that P1 { underwent an early} starburst. \cite{conroy_h3_disk} claim that similar behaviour, at approximately the same metallicity, can be seen in prograde ($L_{\rm z}>-500~{\rm kms}^{-1}$) stars in the stellar halo.

{ If indeed the early starburst hypothesis is correct,}
the star forming gas reservoir had abundances typical of field stars in the halo { at same [Fe/H]}. This starburst must have been {short-lived} enough that either the gas was consumed entirely or star formation was quenched before SNe Ia could make an important contribution to the chemical enrichment of the gas. 


\subsection{The abundance pattern of the    P2 population}\label{pattern_p2}

Unlike their P1 counterparts the stars belonging to the P2 population exhibit, at the low metallicity end ([Fe/H]$\simeq$--1.7) depleted Mg ([Mg/Fe]$\simeq$--0.2), enhanced Si  ([Si/Fe]$\simeq$0.40), and very strongly enhanced Al  ([Al/Fe]$\simeq$0.8). 
Similarly to the case of P1, [Mg/Fe] increases monotonically with [Fe/H].  Conversely, [Si/Fe] initially decreases and flattens to a plateau of [Si/Fe]$\simeq0.3$. [Al/Fe] also increases steeply with [Fe/H], before reaching a peak at [Fe/H]$\simeq-1.4$ and declining towards higher [Fe/H]. 
These abundance patterns are consistent with observations of $\omega$ Cen stars from other groups  \citep[e.g.,][]{jp10,alvarez_garay_wcen}. 

Such abundance patterns are ubiquitous among the Galactic globular clusters that host MPs.
Si and Al enhancement, coupled with Mg depletion, have been postulated to be a clear sign that the star-forming gas reservoir incorporated material processed in stellar interiors by the Mg-Al cycle, during high-temperature ($T\simeq10^7{\rm{K}}$) H-burning in massive as well as AGB stars  \citep[e.g.,][]{arnould_cno_mgal_nena}.
{ As in the case of P1, the steep growth of [Mg/Fe] and [Al/Fe] with [Fe/H] on the low metallicity end suggests the occurrence of an early burst of star formation.  However, [Si/Fe] does not go up with metallicity, making the interpretation of the data for P2 difficult.  This issue is further discussed in  Section~\ref{sec:gce_modelling}.}


\subsection{The abundance pattern of the IM population}\label{pattern_im}

{We discussed the abundance patterns of the P1 and P2 populations in detail in previous sections, concluding that they both differ in substantial ways. 
The IM population is also characterised by substantially different chemistry. One chief difference is the fact that IM stars present a very small dispersion in metallicity, indeed significantly smaller than those of P1 and P2 populations.
The dispersion in the metallicity of the IM population is lower than that of the metal-poor peak of P1 ($\sigma_{\rm{[Fe/H],~IM}}=0.12$, versus $\sigma_{\rm{[Fe/H],~P1}}=0.19$). IM also lacks a tail towards high [Fe/H], a feature that is present in both P1 and P2.
IM also presents a different distribution in the relevant chemical planes. The {light-element} abundance ratios of IM stars lie somewhere between those of P1 and P2 populations at same [Fe/H]. 
For $\upalpha$ elements Mg and Si, IM is closer to P1, whereas for N and Al it shows much larger variance than P1 and P2 stars at the same metallicity.}

Looking more closely, the distribution of the IM population in chemical space is somewhat puzzling{, especially when considering the Al-Fe and N-Fe planes.}
In both planes the trends described by the IM and P2 populations merge seemlessly, with the latter looking like an extension towards higher metallicity of the trends exhibited by the IM population, suggesting a possible chemical evolution link between the two.  
However, the much closer similarity between IM and P1 in the abundances of Si and Mg seem to rule out such a chemical association, arguing instead in favour of a chemical evolution link between those two latter populations.
These contradictory features make it quite difficult for one to devise a clear qualitative evolutionary path connecting these three populations.

\subsection{On the absence of a chemical evolution history connecting the P1, P2, and IM populations}\label{connect_pops}

{The abundance patterns of the three populations displayed in Fig.~\ref{fig:6panel_abundances} are intriguing. 
The fact that P1 and P2 draw widely separate trends in almost all chemical planes suggests no straightforward evolutionary link between these two populations. By the same token, in some chemical planes the IM population seems to be chemically associated with the P1 stars (Mg-Fe and Si-Fe), whereas in others (Al-Fe and N-Fe) there is a hint of a chemical evolution connection between IM and P2 populations.}


Additional insights can be gained by adopting Mg, instead of Fe, as the reference metallicity indicator \citep[which has been previously adopted in works such as][]{mcwilliam08, weinberg_scaling}. 
Unlike Fe, Mg has a single source of enrichment (SNe II), so that the interpretation of Mg abundances is not affected by ambiguities stemming from the enrichment by the ejecta of both SNe II and SNe Ia. 
{However, one of the characteristic abundance patterns of `extreme stars' such as those in our P2 sample is Si-enrichment and Mg-depletion. Those are thought to emerge in stars formed from material processed by high-temperature quiescent hydrogen burning \citep[see][and references therein]{alvarez_garay_wcen}.}

{
The result is shown in Fig.~\ref{fig:6panel_abundances_mg}. It becomes immediately obvious that the three populations are quite detached in critical chemical planes, particularly Al-Mg.  
Stars belonging to the IM population show median [Mg/H]$\simeq$--1.43, whereas the P1 and P2 populations start their chemical evolution at [Mg/H]$\simeq$--1.6 and --2.0, respectively.  
Moreover, P2 stars do not seem to constitute an extension of the IM trend in either the [Al/Mg]-[Mg/H] or the [N/Mg]-[Mg/H] planes.  P2 stars also differ substantially from the P1 and IM in [Si/Mg] on the metal-poor end.  
In these planes, however, IM and P1 populations are similar, though slightly different in [Fe/Mg] and [Si/Mg], while differing quite substantially in [Al/Mg]. 

This exercise demonstrates that the seeming chemical evolution connection between the P2 and IM populations, apparent in the Al-Fe and N-Fe planes falls apart when Mg is taken as the reference metallicity indicator.  
By the same token, while the IM population is more similar to the metal-poor end of the P1 population, differences are large enough that it is not quite easy to conceive how one can evolve from the other.  
We next consider how further examination of the abundance pattern of the IM population can help resolving this conundrum.
}



\subsection{The resemblance of the IM population to mono-metallic Galactic globular clusters} \label{gcs_im}



\begin{figure}
    \includegraphics[width=\columnwidth]{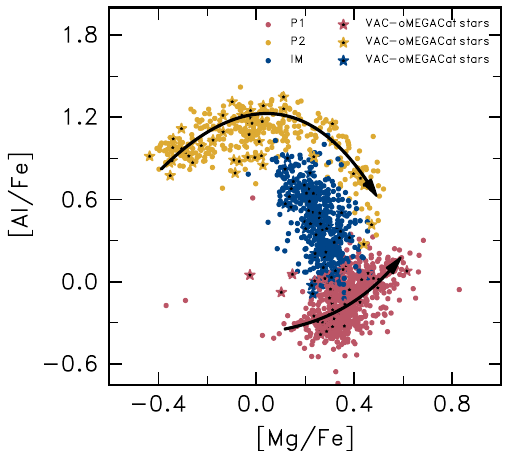}
    \caption{The Mg-Al anticorrelation plotted for the P1 (red), IM (blue), and P2 (yellow) populations in $\omega$ Cen. The black arrows show, approximately, the direction in which [Fe/H] evolves with [Mg/Fe]. It is remarkable that the P2 and IM populations are quite well separated, and that there is little, if any, evolution with [Fe/H] for the IM population. This is one piece of evidence that the IM population may not be the product of {\it in-situ} star formation at all, but rather a `fossil population' formed by the {spiralling in} of a globular cluster into {the centre of the $\omega$ Cen host galaxy}. Stars that appear in both the VAC and oMEGACat catalogues are marked by filled stars, to illustrate that the stars exhibit the same abundance pattern as the sample they are drawn from.}
    \label{fig:mgal}
\end{figure}

Fig. \ref{fig:mgal} shows the distribution of our populations on the Mg-Al plane.  The arrows indicate the direction of [Fe/H] growth in the cases of P1 and P2 (as discussed in Section~\ref{pattern_im}, the IM population has very small dispersion in [Fe/H]). 
Most notably, the IM population shows a significant anticorrelation, at approximately fixed [Fe/H], between [Al/Fe] and [Mg/Fe]. 
This is consistent with the {behaviour of so-called second-generation} populations within {monometallic} Galactic globular clusters \citep[e.g.,][]{carretta_mgal_1851,carretta_mgal_6752,meszaros2015,schiavon2017b,nataf2019,meszaros2020}. 
On this basis, we speculate that the IM population in fact {consists of between one and a few metal-poor field globular clusters that spiralled into the core of the host galaxy that $\omega$~Cen used to be the NSC of.}

To test this hypothesis, we search the \cite{vac_paper} VAC for globular clusters whose median [Fe/H] lies within $\pm\sigma_{{\rm{[Fe/H]}}}$ of the median value of the IM population (Table~\ref{tab:mgal_properties}). For those Galactic globular clusters, we coarsely select anomalous stars by imposing the criterion that they have [Al/Fe] above the line where ${\rm{[Al/Fe]}}=m*{[\rm{Fe/H}]}+c$ with $m=0.5$ and $c=-0.3$.  We summarise their properties in Table \ref{tab:mgal_properties}.

\begin{table}
    \centering
    \begin{tabular}{c c c|}
        \toprule
        GC name & Median [Fe/H] & $\sigma_{\rm [Fe/H]_{2P}}$\\
        \midrule
        NGC 4147 & -1.63 & 0.06\\
        NGC 5466 & -1.81 & 0.09\\
        NGC 5634 & -1.72 & 0.06\\
        NGC 6093 & -1.61 & 0.004\\
        NGC 6144 & -1.80 & 0.00\\
        NGC 6273 & -1.71 & 0.14\\
        NGC 6656 & -1.70 & 0.10\\
        NGC 6809 & -1.76 & 0.08\\
        Terzan 10 & -1.62 & 0.10\\
    \end{tabular}
    \caption{Globular clusters in the APOGEE VAC whose median P2 [Fe/H] lies within $\pm\sigma_{\rm [Fe/H]_{IM}}$.}
    \label{tab:mgal_properties}
\end{table}

\begin{figure}
    \includegraphics[width=\columnwidth]{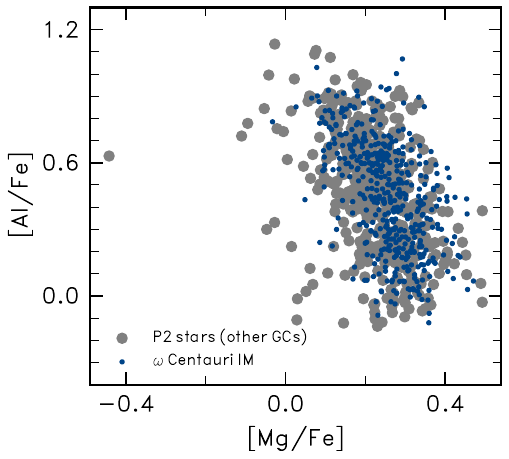}
    \caption{The Mg-Al anticorrelation of the IM stars identified in $\omega$ Cen, and those in other Galactic globular clusters present in the VAC whose properties are summarised in Table \ref{tab:mgal_properties}, having been identified as having median [Fe/H] within $\pm\sigma_{[{\rm{Fe/H},~IM}]}$ of the IM population. The fact that these stars all show a distribution in this plane consistent with the IM population lends credence to the scenario whereby $\omega$ Cen experienced spiralling in of field globular clusters from its host galaxy during distant cosmic epochs.}
    \label{fig:mgal_comparison}
\end{figure}

Fig. \ref{fig:mgal_comparison} shows the GC stars identified by the above criterion on the Mg-Al plane ({gray} points) overlaid onto stars comprising the $\omega$ Cen IM population (black points). 
The agreement is remarkable, in support of our hypothesis.
Assuming that the onset of the MP phenomenon followed the formation of P1, it makes little sense that they would form with such an abundance pattern after the starburst.


In the Section~\ref{apo_chm}, we map the stars comprising our P1, P2, and IM sample onto the oMEGACat chromosome map, which we derived in $\S$\ref{subsubsec_chm}.

\subsection{Cerium abundances of $\omega$~Cen stars} \label{sec:cerium}

\begin{figure*}
    \includegraphics[width=\textwidth]{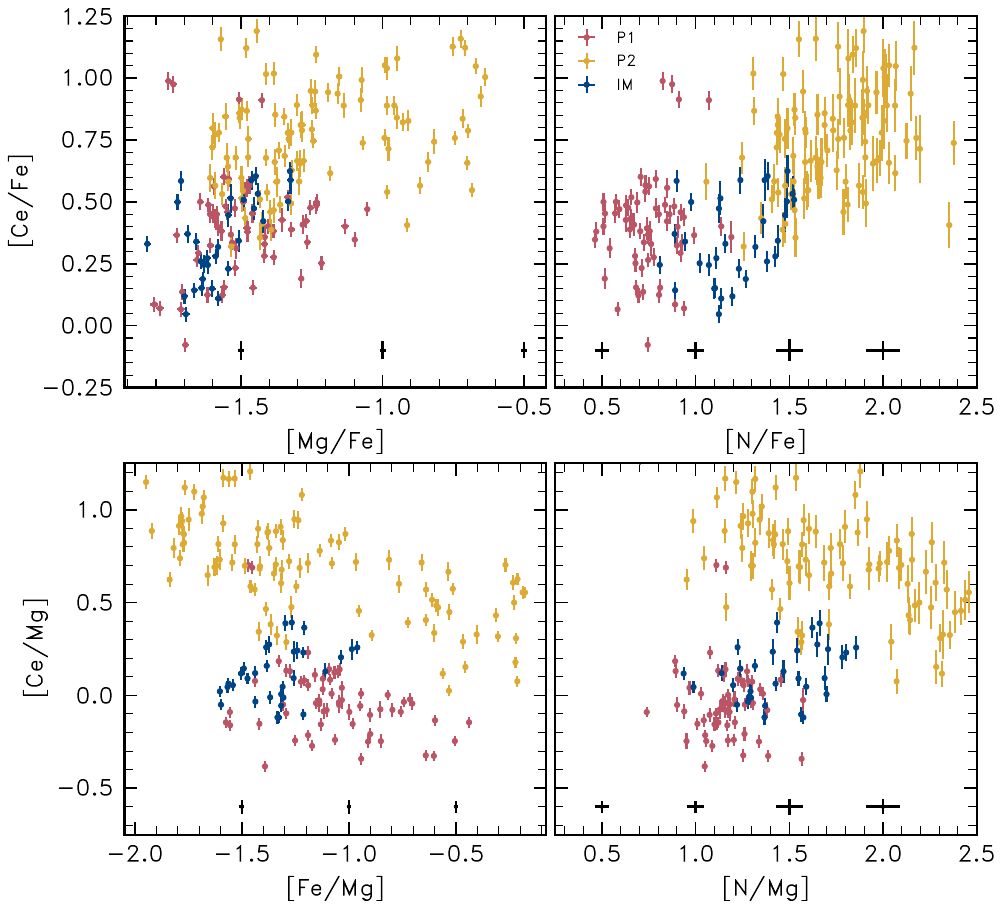}
    \caption{Ce abundances from the BAWLAS catalogue for $\omega$ Centauri stars present in both that catalogue and the VAC. The average BAWLAS-derived uncertainties are illustrated by black crosses in 0.5 dex bins at the bottom right of each panel.}
    \label{fig:bawlas}
\end{figure*}

In this Section we discuss the s-process element abundance patterns of our three populations by supplementing our VAC sample with re-derived abundances included in the BACCHUS \citep{masseron16_bacchus} Analysis of Weak Lines in APOGEE Spectra (BAWLAS) \citep{hayes2022} value-added catalogue. 
{In Fig. \ref{fig:bawlas} $\omega$~Cen stars are displayed on four planes: [Ce/X] vs. [X/H] ({\it left panels}), and [Ce/X] vs. [N/X] ({\it right panels}), where X=Fe on the top panels and X=Mg on the bottom panels. We first focus on the top panels.  As pointed out by \cite{milone_atlas_chm}, $\omega$~Cen displays a type-II GC behaviour, whereby it contains stars with a range of [Ce/Fe] abundance ratios, which in turn are correlated with [Fe/H].  We note however that this correlation is only quite strong within the P2 population.  The P2 population is characterised by higher [Ce/Fe] values, on average, than both the IM and P1 populations, which in turn have very similar [Ce/Fe]. We also note that there is a very strong correlation between [Ce/Fe] and [N/Fe].}

{On the bottom left panel one can see the dependence of [Ce/Mg] on [Mg/H].  In the case of the P2 population, [Ce/Mg] decreases strongly towards higher [Mg/H], whereas is is roughly constant in P1 stars. 
In the IM population [Ce/Mg] shows a large scatter at relatively constant [Mg/H]. This behaviour resembles  that displayed by our sample in the [Al/Mg] vs. [Mg/H] plane, although with larger scatter, presumably due to higher uncertainties in the Ce abundances.  
This result suggests that Ce and Al share a common nucleosynthetic source in $\omega$~Cen.  Finally, the bottom right panel shows that the strong correlation between [Ce/Fe] and [N/Fe] disappears when Fe is replaced by Mg as the metallicity indicator.  
In fact, the P2 population shows an anti-correlation between [Ce/Mg] and [N/Mg].  The latter trend switch is caused by the steep correlation between [Mg/Fe] and [Fe/H] in P2, indicating that this population is enriched in Mg at a faster pace than Fe, which is consistent with a starbust behaviour, as discussed in Section~\ref{sec:gce_modelling}.

}

{In Section~\ref{apo_chm} we discuss the possible connections between the Ce abundances in $\omega$~Cen stars and their distribution on the chromosome map.}


\subsection{APOGEE stars on the chromosome map} \label{apo_chm}

\begin{figure*}
    \includegraphics[width=\textwidth]{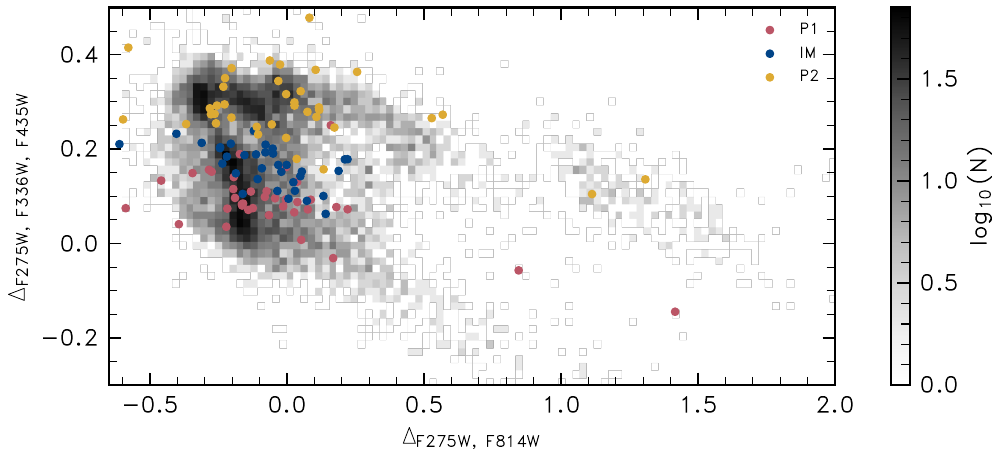}
    \caption{ The chromosome map for $\omega$ Cen, constructed using multi-band photometry from the oMEGACat catalogue, represented as a 2d histogram where in 0.05 dex bins of $\Delta_{\rm F275W,~F814W}$ amd $\Delta_{\rm F275W,~F336W,~F435W}$ each pixel shows the number of stars contained in each bin. Coloured points indicate the positions of the P1 (red), P2 (yellow), and IM (blue) stars that overlap between the VAC and oMEGACat.}
    \label{fig:chm}
\end{figure*}

In this section, we examine how the abundance patterns exhibited by the three populations identified in our study map into their loci on the so-called `chromosome map' (ChM). 
Fig. \ref{fig:chm} shows the ChM derived from oMEGACat, as described in $\S$\ref{subsubsec_chm}. Data from oMEGACat are represented as a 2d histogram where each pixel {is shaded according to the} (logarithmic) number of stars in 0.05~mag bins in $\Delta_{\rm F275W,~F336W,~F435W}$ and $\Delta_{\rm F275W,~F814W}$. Adopting the same colour scheme as in Figs. \ref{fig:1_mdf}-\ref{fig:mgal_comparison}, coloured points show the 134 giants {in common between} oMEGACat and APOGEE.

Fig. \ref{fig:chm} shows that the P1, P2, and IM stars occupy separate sequences on the ChM as they do in Figs. \ref{fig:6panel_abundances}, \ref{fig:6panel_abundances_mg}, and \ref{fig:mgal}. 
The three sequences approximately correspond to sequences that are clearly distinguishable in the oMEGACat data, running diagonally from the top left to the bottom right of the plane.
The spread of the data in this direction is associated with the metallicity variation within each population, in the sense that [Fe/H] grows towards the bottom right corner of the pseudo-colour-colour plane \citep[e.g.,][]{milone_atlas_chm}.  
The three sequences are also vertically displaced.  As discussed previously \citep[e.g.,][]{milone_atlas_chm}, vertical shifts are associated with variations in the abundances of light elements.
Thus, the positions of the P1, IM, and P2 populations on this plane are consistent with the light element abundance patterns displayed in Figures~\ref{fig:6panel_abundances} and \ref{fig:6panel_abundances_mg}.

The P2 population overlaps a locus that encompasses the overdensity at the top left of the plot ($\Delta_{\rm F275W,~F336W,~F435W}\simeq-0.3$), extending along a ``plume'' that runs diagonally towards redder $\Delta_{\rm F275W,~F814W}$ and bluer $\Delta_{\rm F275W,~F336W,~F435W}$. 
Conversely, the P1 population is associated with the bluest $\Delta_{\rm F275W,~F336W,~F435W}$ sequence, although its most metal-rich stars do not seem aligned with it on the red $\Delta_{\rm F275W,~F814W}$ end.
Finally, the IM population occupies a locus at $\Delta_{\rm F275W,~F336W,~F435W}\simeq0.15$, located in between the regions populated by P1 and P2, with a much shorter range in $\Delta_{\rm F275W,~F814W}$ than P1 and P2. 

{Fig.~\ref{fig:chm} is very revealing.}  The chemical complexity of $\omega$~Cen has been discussed by various groups \citep[e.g.,][]{jp10, milone_atlas_chm, marino_milone_chm_tagging, alvarez_garay_wcen}.  
In particular, studies have led to reports that there may be as many as 15 stellar populations in $\omega$~Cen on the basis of its metallicity distribution function and light element abundance patterns \citep[e.g.,][]{pancino2000, sollima_mps}.  
According to that interpretation of the data, such stellar population complexity manifests itself in the ChM of  Fig.~\ref{fig:chm} 
in the form of multiple density peaks, each associated with a stellar population of different metallicity and light-element abundance pattern.

Figs.~\ref{fig:6panel_abundances}, \ref{fig:6panel_abundances_mg} and~\ref{fig:chm} suggest that the stellar population mix of $\omega$~Cen is in fact much simpler than previous studies have suggested. As discussed above, metallicity explains the diagonal extension of each of the three sequences, whereby $\Delta_{\rm F275W,~F336W,~F435W}$ decreases as a function of $\Delta_{\rm F275W,~F814W}$.  Interestingly, P1 and P2 extend over a much wider range in $\Delta_{\rm F275W,~F814W}$ than IM, as one would expect from the MDFs of these populations as well as their distributions in multiple chemical planes.

It is noteworthy that the P1 and P2 sequences run in parallel on the ChM, while being widely separated in the chemical planes displayed in Figs.~\ref{fig:6panel_abundances}-\ref{fig:mgal}, where they display marked differences in their abundance patterns. Both of them straddle a wide range in colour, containing multiple density peaks that are clearly visible in the oMEGACat data. 
One example is the overdensity at $\Delta_{\rm F275W,~F814W}\simeq1.25$, which corresponds to the high metallicity end of P2 at [Fe/H]~$\simeq -0.6$.  This metal-rich component of P2 can also be easily identified in the CMD and Kiel diagram of Fig.~\ref{fig:cmds}. 

The distribution of APOGEE stars on the ChM in Fig.~\ref{fig:chm} and on chemical planes in Figs.~\ref{fig:6panel_abundances} and \ref{fig:6panel_abundances_mg} thus strongly suggests that {\it the multiple densities seen in the ChM are not independent stellar populations}.  Instead, each one of them is connected to one out of three populations characterised by a particular set of light-element abundances and covering different ranges in metallicity. 
These seemingly detached substructures occurring along either of the three parallel sequences are in fact connected by a common abundance pattern, while differing chiefly in terms of overall metallicity. 
As discussed in Section~\ref{sec:gce_modelling}, this behaviour is consistent with all three populations having formed from detached star formation episodes operating on distinct gas reservoirs. 
Under this interpretation, the multiple density peaks in the oMEGACat data for the P1 and P2 populations, represent metallicity peaks associated with a bursty star formation history.  They are not apparent in the APOGEE MDF because, unlike oMEGACat, the APOGEE data do not sample the populations of $\omega$~Cen densely enough to resolve those peaks.  

{Finally, it has been pointed out that the distribution of GC stellar populations on the ChM are determined by variations in light-element abundance patterns (associated with the multiple populations phenomenon) and metallicity (in the case of type II GCs). 
Indeed the loci of chemically selected $\omega$~Cen stars on the ChM agree well with that interpretation of the data, as the diagonal stretch of each of the three populations in the ChM correlates with` their ranges in metallicity, whereas
differences in light element abundance account for their relative vertical displacements.  One additional component must be considered, though.  \cite{milone_atlas_chm} have shown that stars enhanced in s-process elements in type II GCs extend over a branch that is located to the red of the main diagonal branch on the ChM.  The size of the BAWLAS subsample in common with oMEGACat is unfortunately not large enough for a robust  confirmation of that trend.  Nevertheless, the fact that the P2 population alone is almost entirely responsible for the variance in [Ce/Fe,Mg] may help explain the apparent vertical scatter in its associated branch in the ChM.}



\subsection{Summary of the observational evidence} \label{summary_obs}

We have examined the chemical properties of each subpopulation identified on the basis of their abundances using $k$-means clustering. 
There are two populations (P1 and P2) that seem to exhibit { tight correlations between abundance ratios and metallicity, which are characteristic of chemical enrichment during periods of extended star formation.} 
Conversely, there is a population characterised by its comparative lack of an [Fe/H] spread which is consistent with that measured in other Galactic GCs present in the VAC.  
This population (IM) has light element abundance ratios that are intermediate to those of the P1 and P2 populations at same metallicity. Most importantly, it exhibits the well-known Mg-Al anticorrelation which strongly resembles that of other GCs. 

We interpret these observations as evidence that $\omega$~Cen can broadly be described as being comprised of three populations, which formed in separately.  One of them (IM) {also} shows an abundance pattern at fixed [Fe/H] that is consistent with that of monometallic metal-poor Galactic GCs, suggesting that it may result from the inspiralling of at least one GC into the central potential of $\omega$~Cen's host galaxy.
Conversely, the other populations (P1 and P2) { are characterised by metallicity spreads and tight correlations between metallicity and abundance ratios, which suggest that their chemical evolution was influenced by an early burst of star formation}.
Because the abundance ratios of these two populations differ significantly at every metallicity these bursts of star formation have likely occurred at different points in space, time, or both.


In $\S$\ref{sec:gce_modelling} we rely on models of galactic chemical evolution to speculate on the histories of gas infall and star formation that could be responsible for the chemical properties of the P1 and P2 populations. 
Following from our interpretation of its abundance patterns in $\S$\ref{sec:chemistry}, we assume that the IM population represents {a combination of typical mono-metallic GCs.}
In view of the lingering uncertainties regarding the origin of such anomalies \citep{renzini2015,bastian_gc_review}, we refrain from modelling the chemical evolution of the IM population. Instead, we focus on the P1 and P2 populations.  
In $\S$\ref{sec:discussion} we discuss  scenarios that may explain the co-existence of these three populations today within the $\omega$~Cen stellar system.

\section{Modelling the Chemical Evolution of \texorpdfstring{$\omega$}~Cen}
\label{sec:gce_modelling}

We begin this section by describing the prescriptions adopted for the key ingredients of our models. Following that we compare our predictions with the chemical composition data available for the P1 and P2 populations. All our calculations are based on the {\tt Versatile Integrator for Chemical Evolution (VICE)} galaxy chemical evolution modelling code \citep{vice1, vice2, vice3}.

\subsection{Model prescriptions}

In $\S$\ref{sec:chemistry} we speculate that the abundance patterns shown by the P1 and P2 populations on the Al-Fe and $\upalpha$-Fe planes are consistent with chemical evolution resulting from  bursts of star formation. Such  behaviour has been seen identified in data for several Local Group dwarfs such as the Large Magellanic Cloud \citep{nidever_lmc}, Sagittarius dSph \citep{hasselquist_sgr}, and Fornax \citep{hasselquist_lgds,fernandes2023}. 
In GCE models, bursts of star formation can be brought about by invoking either a sudden inflow of gas, or by an enhancement of the star formation efficiency (SFE). 
This follows from the fact that both kinds of event can significantly enhance the SFR, and thus temporarily enhance the instantaneous metal contribution by CCSNe such that it {exceeds} that of SNe Ia before {steadily converging} toward an equilibrium abundance \citep[see][for a thorough discussion of $\upalpha$-enhancement due to sudden star formation events]{weinberg_sudden_events_17}.

Thus, for P1 and P2 we require prescriptions in VICE that can reproduce the trend of increasing [$\upalpha$/Fe] as a function of [Fe/H]. 
In the case of P1, the characteristic `rising' behaviour is similar to that observed in the Milky Way thick disk in the H3 and APOGEE surveys presented in \cite{conroy_h3_disk}.  
We speculate that this indicates the occurrence of a period of initially inefficient star formation that was followed by a sudden enhancement of the SFE to form the rise in [$\upalpha$/Fe]. 
\cite{conroy_h3_disk} claim that the burst of star formation coincided with the formation of the high-$\upalpha$ disk from an initially kinematically hot population. 
However, to caveat this picture we point out that \cite{chen_alpha_rise} found that such a change in the SFE was not necessary to produce {the observed} enhancement of [$\upalpha$/Fe] and that the inflow of fresh gas (i.e.,`cold mode' accretion) can also produce this behaviour.

Firstly, we adopt the widely-used linear-exponential form of the gas inflow rate as a function of cosmic time, $t$, by the following equation:
\begin{equation}
    \label{eq:inflow}
    \dot{M_{{\rm in}}} = \frac{M_{{\rm i}}}{\tau_{{\rm in}}}\frac{t}{\tau_{{\rm in}}} {\rm{e}}^{\frac{-t}{\tau_{{\rm in}}}},
\end{equation}
where $M_{{\rm i}}$ is the inflow mass scaling factor in units $\rm{M}_\odot$, and $\tau_{in}$ is the e-folding timescale in Gyr. We describe the SFE in terms of its inverse, the gas consumption timescale ($t_{{\rm g}}$), and assume it takes the form of a sigmoid function (as used in \citealp{mason_knee}) such that:
\begin{equation}
    \label{eq:tg}
    t_{{\rm g}}(t) = t_{\rm g,b} + \frac{t_{\rm g,i}}{1+\exp{[-k(t-t_b)}]},
\end{equation}
where $t_{\rm g,i}$ and $t_{\rm g,b}$ are respectively initial and the final values of $t_{\rm g}$, $k$ is the multiplicative factor of the exponent of the sigmoid, and $t_{b}$ is the time at which the SFE begins to {increase}.


We assume that $\omega$ Cen's P1 and P2 populations formed such that no gas enriched by either population was mixed into the other, and thus attempt to fit two separate open box single-zone models. We assume that the amount of gas removed by feedback at any given timestep ($t_{\rm step}$) within this box is given by:
\begin{equation}
    \label{eq:outflows}
    M_{{\rm out},~t=t_{\rm step}} = \eta~{\rm SFR}(t=t_{\rm step}),
\end{equation}
where {SFR is the star formation rate in M$_\odot$~Gyr$^{-1}$}, $\eta$ is the outflow mass loading factor in units ${\rm Gyr}$, and $t_{\rm step}$ is the time corresponding to a given timestep in Gyr. 
Furthermore, we assume that the inflowing gas is {\it not} of a primordial composition whose abundances were set by Big Bang nucleosynthesis. {Instead, in both cases we assume that the chemical composition of the inflowing gas matched that of stars at the low end of the [Fe/H] distribution.
Interestingly, in the case of P1, that happens to be similar to the chemical composition of the Galactic halo at the same [Fe/H].
}

We follow the procedure outlined in \cite{johnson_mcmc} in order to infer best-fitting models to the APOGEE data for both P1 and P2 stars, adopting the above prescriptions for the metallicity of the inflowing gas, the outflow mass loading factor, the gas consumption timescale, and the history of gas inflow. 
There are nine free parameters in our model, given by $\theta=[t_{\rm g,i},~k,~t_{\rm g,b},~t_{b},~\eta,~\tau_{\rm in},~t_{\rm tot},~Z_{\rm in,~Fe},~Z_{\rm in,~Mg}]$. Parameters $L$, $t_0$, $k$, $t_{\rm g,0}$, $\nu$, and $\tau_{\rm in}$ are defined in Eqs. \ref{eq:inflow}-\ref{eq:outflows}.
The remaining parameters $t_{\rm tot}$, $Z_{\rm in,~Fe}$, and $Z_{\rm in,~Mg}$ are the total cosmic runtime of the model in Gyr, and the inflowing abundances of Fe and Mg into the box, respectively.   

We assume flat, uniform priors on each model parameter with additional conditions that, for P1, ensure:
\begin{enumerate}
    \item $0\leq \eta<100$ Gyr
    \item $t_{\rm tot}<t_{\rm cosmo}$
    \item $Z_{\rm in,~Mg}>Z_{\rm in,~Fe}$,
\end{enumerate}
where $t_{\rm cosmo}$ is the {age of the universe}, taken as the \cite{planck_latest} value of $t_{\rm cosmo}\simeq 13.8~{\rm Gyr}$ (see Table 2 of that paper). 
For P2, we enforce identical conditions except for {\it (iii)} where we enforce $Z_{\rm in,~Mg}<Z_{\rm in,~Fe}$, drawing from a uniform priors in the ranges $Z_{\rm in,~Mg}=10^{-6}\times1-10$ and $Z_{\rm in,~Mg}=10^{-5}\times1-10$. 
Model parameters are summarised in tables ~\ref{tab:p1_summary} and \ref{tab:p2_summary}.

\subsection{Modelling results}

{\subsubsection{Initial caveats}}
\label{sec:caveats}

When fitting models with large sets of free parameters, it is good practice to impose physically meaningful constraints on the range of values parameters should span. It is common practice in applications of GCE models to maximise the constraints by considering both the run of abundance ratios of stellar populations with metallicity and the probability density of observing a star at a given metallicity (the Metallicity Distribution Function, MDF).


The case of $\omega$~Cen is not simple in that regard.  As the likely former nuclear cluster of a satellite of the Milky Way that was accreted many Gyr ago \citep{massari_gc, limberg_sausage}, $\omega$~Cen must have been subject to strong tidal stripping. This has been confirmed by studies reporting detection of $\omega$~Cen stars in the halo \citep{ibata_fimbulthul_wcen, galah_fimbulthul}.
There are at least two important implications of that fact for our chemical evolution models of $\omega$~Cen.  First, stars of varying age and chemical composition may have been stripped over the lifetime of $\omega$~Cen, so that the MDF of the surviving stellar population may not reflect its history of star formation and chemical enrichment.  
Second, the possible occurrence of gas loss due to tidal stripping means that gas removal cannot be assumed to originate purely from stellar feedback, as implied by Eq.~\ref{eq:outflows}.

To address these issues, we fit two sets of models to the data, in which the MDF is or is not used as a constraint.  In this way, we can evaluate whether there is consistency between the evolution on the abundance ratios and the bulk chemical enrichment of the system.  Important discrepancies in the results obtained in the two model fits could lend insights into the history of tidal stripping of the system.

Another important warning must be brought to the reader's attention at this stage.  The MDFs for each population displayed in Fig.~\ref{fig:1_mdf} are based on  samples of several hundred stars, spread over a range of over a decade in [Fe/H] {so that, as pointed out in Section~\ref{subsubsec_chm}, they lack the resolution needed to detect the multiple peaks in $\omega$~Cen's real MDF.} 
Such a limitation has an obvious impact on our ability to discern sharp time variations in $\omega$~Cen's star formation rate, likely caused by tidal interactions as its host galaxy collapsed under the gravity of the Milky Way halo. 
Furthermore, our sample is largely limited to bright and isolated giants and we do not take into account the selection function of APOGEE in our analyses.

While definitely not fine-grained, the SFHs inferred from our modelling should nonetheless be able to account for the broad distribution of $\omega$~Cen's stellar populations in chemical space,  shedding light on the evolutionary history leading up to its current state.

\subsubsection{``MDF-Constrained Models''} \label{sec:constrained}

\begin{table}
    \centering
    \begin{tabular}{ ccc }
        \toprule
          & Parameter & Value \\
        \midrule
        SFE & $t_{\rm g, i}~[{\rm Gyr}]$ & $172\pm^{7.4}_{5.7}$ ~ $\mathbf{139\pm^{5}_{6}}$ \\[1mm]
         & $t_{\rm g,b}~[{\rm Gyr}]$ & $1.01\pm^{0.17}_{0.10}$ ~ $\mathbf{1.08\pm^{0.04}_{0.06}}$ \\[1mm]
         & $t_{\rm b}~[{\rm Gyr}]$ & $3.94\pm^{0.03}_{0.05}$ ~ $\mathbf{3.49\pm^{0.10}_{0.13}}$ \\[1mm]
         & $k$ & $17.8\pm^{0.6}_{0.31}$ ~ $\mathbf{12.4\pm^{0.57}_{0.41}}$\\[1mm]
        \midrule
        Inflows & $t_{\rm in}$ [Gyr] & $2.11\pm^{0.24}_{-0.26}$ ~ $\mathbf{3.44\pm^{0.31}_{0.32}}$\\[1mm]
        & $Z_{\rm in,~Mg}$ & $9.72\pm0.02\times10^{-6}$ ~ $\mathbf{1.19\pm^{0.02}_{0.02}\times10^{-5}}$\\[1mm]
        & $Z_{\rm in,~Fe}$ & $9.47\pm^{0.02}\times10^{-6}$~ $\mathbf{1.21\pm^{0.02}_{0.02}\times10^{-5}}$\\[1mm]
        &$\eta~[{\rm Gyr^{-1}}]$ & $67.1\pm{0.9}$ ~ $\mathbf{33.5\pm^{1.3}_{1.19}}$\\[1mm]
        \midrule
        Other & $t_{\rm total}$ [Gyr] & $4.80\pm^{0.36}_{0.32}$~ $\mathbf{4.38\pm^{0.11}_{0.12}}$ \\
        \bottomrule
    \end{tabular}
    \caption{A table showing the median values of the posterior probability distributions of the model parameters for our P1 fits. Uncertainties are taken as the interquartile range of the posterior PDF, with bold values indicating models with relaxed constraints on the SFH.}
    \label{tab:p1_summary}
\end{table}

\begin{figure}
\includegraphics[width=\columnwidth]{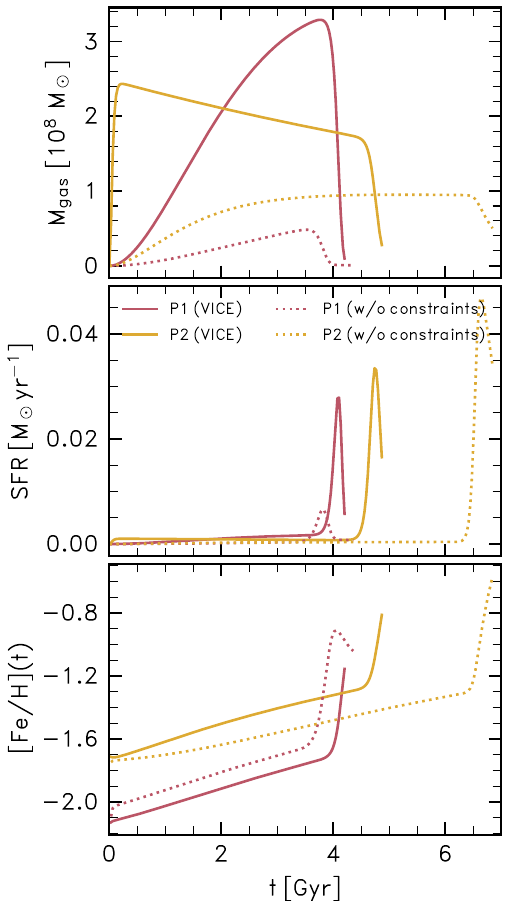}
    \caption{Histories of gas infall, star formation, and Fe-evolution with cosmic time from the models corresponding to the median parameters drawn from the posterior PDFs of our model fits to the P1 and P2 samples, also seen in Figs. \ref{fig:alpha_fe_p1} and \ref{fig:alpha_fe_p2}.}
    \label{fig:sfrs_p1_p2}
\end{figure}

\begin{figure*}
    \includegraphics[width=\textwidth]{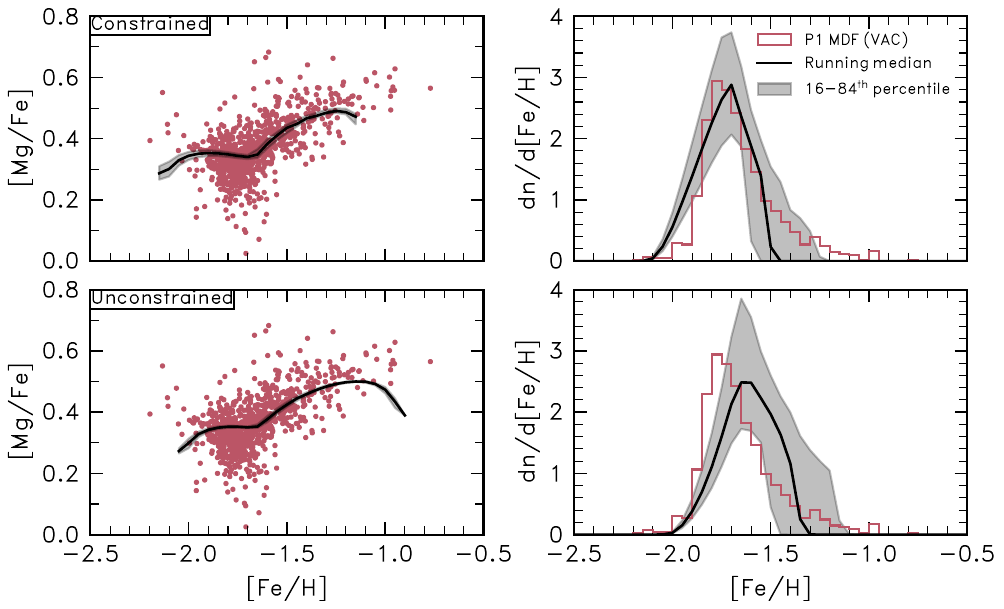}
    \caption{MDFs and Mg-Fe planes from the model fits to the P1 stars, where black solid lines correspond to the model corresponding to the median of the posterior PDFs. In each case, MDFs generated by the model have been convolved with the median uncertainty in [Fe/H] in the APOGEE VAC for P1 stars in $\omega$ Cen. The top and bottom rows correspond to the models described in $\S$\ref{sec:constrained} and $\S$\ref{sec:unconstrained}, respectively. Grey shaded regions indicate the range of MDFs and abundance tracks produced by randomly sampling model parameters between the 16$^{\rm th}$ and 84$^{\rm th}$ percentiles of the posterior probability distributions. The precipitous decline in the probability density of stars with [Fe/H]$>-1.7$ reflects the rapid consumption of the star-forming gas following the onset of the starburst.}
    \label{fig:alpha_fe_p1}
\end{figure*}

We start by studying models optimised to match both the distribution of stars on the Mg-Fe plane and the MDFs of the two populations. {The resulting histories of gas infall and star formation, as well as the evolution of [Fe/H] with time, are shown as solid lines in Fig.~\ref{fig:sfrs_p1_p2}.}  
The model fit to the P1 stars favours parameters with initially inefficient star formation, characterised by a gas consumption timescale $t_g\simeq227$ Gyr. 
At $t\simeq4.49~{\rm Gyr}$, $t_g$ declines precipitously to $\simeq1.16~{\rm Gyr}$, resulting in a starburst in the gas reservoir. Gas of a halo-like composition, such that ${\rm [Mg/Fe]_{\rm in}}=0.35$, flows into the box over a timescale of $t_{\rm in}=0.40~[{\rm Gyr}]$. 
Outflows are highly efficient in removing gas per unit star formation, such that $\eta=38~{\rm Gyr}$. The top panels of Fig.~\ref{fig:alpha_fe_p1} shows the predicted model track of [Mg/Fe]$(t)$ vs. [Fe/H]$(t)$ and MDFs for the P1 model, plotted against the P1 stars identified in $\S$\ref{sec:chemistry}. 
{The starburst causes an increase in the frequency of SNe~II, and a boost in $\alpha$-element production, which manifests itself in the form of a sudden change in the slope of the model in the Mg-Fe plane at ${\rm [Fe/H]\approx-1.65}$.}

The top panels of Fig. \ref{fig:alpha_fe_p1} shows the results of our constrained models for P1. The left panel shows the observed Mg-Fe plane, with a solid line indicating the model corresponding to the median parameters of the posterior probability distribution produced by the fitting. 
The right panel shows the observed MDF as a histogram in 0.05 dex bins of [Fe/H]. In both panels, the solid line corresponds to the model prediction generated for the median parameters of the posterior probability distribution. 
The shaded region indicates 1,000 realisations of models with parameters randomly drawn, with values ranging between the 16$^{\rm th}$ and 84$^{\rm th}$ percentiles of the posterior PDF. 
Overall, a good match to the data is achieved. The normalisation and shape of the MDF is reproduced, along with the [Fe/H] corresponding to the peak of the MDF. However, the model under-predicts the number of metal-rich stars.

\begin{table}
    \centering
    \label{tab:p2_summary}
    \begin{tabular}{ ccc }
        \toprule
          & Parameter & Value \\
        \midrule
        SFE & $t_{\rm g,i}~[{\rm Gyr}]$ & $235\pm^{5}_{6}$ ~ $\mathbf{269\pm^{98}_{82}}$\\[1mm]
         & $t_{\rm g,b}~[{\rm Gyr}]$ & $1.51\pm^{0.06}_{0.09}$~ $\mathbf{1.48\pm^{0.12}_{0.11}}$ \\[1mm]
         & $t_{\rm b}~[{\rm Gyr}]$ & $4.43\pm^{0.71}_{0.92}$~ $\mathbf{6.31\pm^{0.11}_{0.08}}$ \\[1mm]
         & $k$ & $16.5\pm^{0.8}_{0.6}$~ $\mathbf{22.8\pm^{2.6}_{4.7}}$ \\[1mm]
        \midrule
        Inflows & $t_{\rm in}$ [Gyr] & $0.3\pm^{0.01}_{0.01}$ ~ $\mathbf{0.77\pm^{1.29}_{0.71}}$\\[1mm]
        & $Z_{\rm in,~Mg}$ & $1.0\pm0.06\times10^{-6}$~ $\mathbf{2.35\pm^{0.04}_{0.05}\times10^{-5}}$\\[1mm]
        & $Z_{\rm in,~Fe}$ & $2.46\pm^{0.07}_{0.04}\times10^{-5}$~ $\mathbf{3.60\pm^{1.30}_{1.68}\times10^{-7}}$\\[1mm]
        & $\eta~[{\rm Gyr^{-1}}]$ & $18.8\pm{0.68}$~ $\mathbf{2.31\pm^{0.77}_{0.10}}$ \\[1mm]
        \midrule
        Other & $t_{\rm total}$ [Gyr] & $4.89\pm{0.06}$ ~ $\mathbf{6.84\pm^{0.08}_{0.06}}$ \\
        \bottomrule
    \end{tabular}
    \caption{A table showing the median values of the posterior probability distributions of the model parameters for our P2 fit. Uncertainties are taken as the interquartile range of the posterior PDF. Bold values indicate models with relaxed constraints on the SFH.}
\end{table}

\begin{figure*}
    \includegraphics[width=\textwidth]{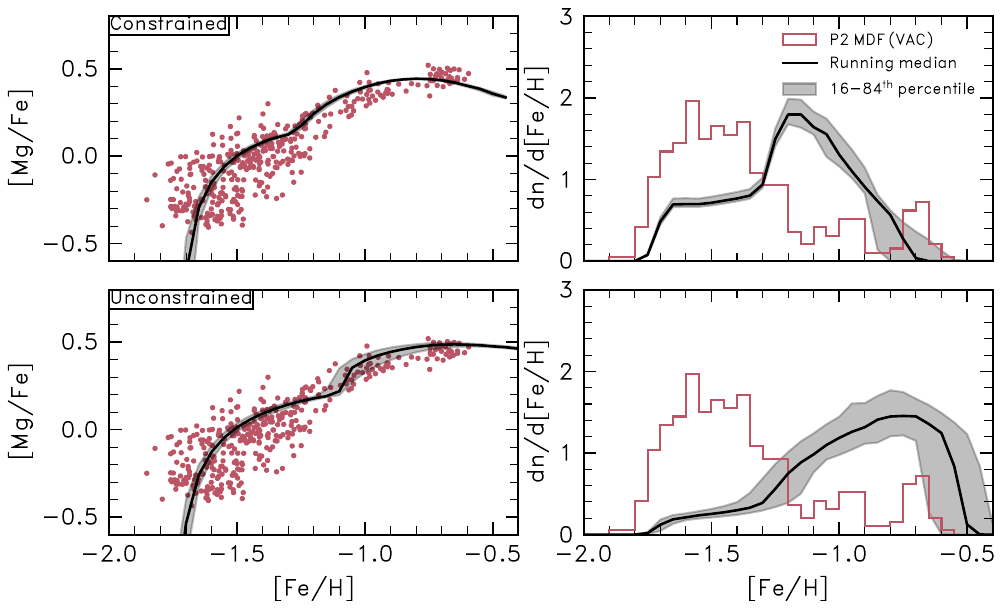}
    \caption{MDFs and Mg-Fe planes from the model fits to the p2 stars, where black solid lines correspond to the model corresponding to the median of the posterior PDFs. In each case, MDFs generated by the model have been convolved with the median uncertainty in [Fe/H] in the APOGEE VAC for P2 stars in $\omega$ Cen. The top and bottom rows correspond to the models described in $\S$\ref{sec:constrained} and $\S$\ref{sec:unconstrained}, respectively. Grey shaded regions indicate the range of MDFs and abundance tracks produced by randomly sampling model parameters between the 16$^{\rm th}$ and 84$^{\rm th}$ percentiles of the posterior probability distributions. Both models over-predict the abundance of metal-rich ([Fe/H]$>-1.2$) stars to varying degrees, in spite of successfully reproducing the abundance pattern of P2. This over-abundance can be attributed to the prescription of a continuous SFH - if there was a period of less-intense star formation during which gas was stripped from the cluster, followed by a fresh round of more intense star formation from the diminished gas reservoir, this could explain the relative paucity of metal-rich stars, and why the model is unsuccessful. Alternatively, these stars could have been stripped during interactions with the Milky Way.}
    \label{fig:alpha_fe_p2}
\end{figure*}

The top panels of Fig.~\ref{fig:alpha_fe_p2}  show comparisons between data and models for the P2 population.
In this case the abundance patterns necessitate the inflowing gas to have abundances characteristic of the so-called `extreme' populations seen in some Galactic globular clusters, such that $[{\rm Mg/Fe}]_{\rm initial}\simeq-0.60$. 
We do not include prescriptions for any of the purported progenitors responsible for these abundance patterns, and instead assume that they have already contributed their metals to the gas reservoir. 
Subsequent chemical evolution from this initial abundance pattern, is driven by enrichment coming from combination of SNe~II, SNe~Ia, and AGB stars.

{As in the case of P1, the P2 population is characterised initially by a very low SFE ($t_{g,i}=235$~Gyr).  Because of a much more vigorous initial gas infall, the star formation rate is higher than in the case of P1.  
About 4.4~Gyr after the beginning of star formation, the SFE surges precipitously ($t_{g,b}=1.51$~Gyr), bringing about a burst of star formation.  Due to the initially higher star formation rate, when the starburst takes place the gas has achieved higher metallicity ${\rm ([Fe/H]\approx-1.3}$).  
As in the case of P1, outflows are also responsible for the termination of star formation.}

The top left panel of Fig. \ref{fig:alpha_fe_p2} shows that the model {matches the distribution of P2 stars on the Mg-Fe plane quite well. 
However, the top right panel shows that the model} significantly over-predicts the number of stars formed from the starburst. Our modelling sugests that these stars form over the course of a continuous history of star formation, and not a bursty one where there can be periods of quiescence or even temporary quenching of star formation, during which time gas mass can be lost due to sources of feedback and stripping. This assumption may be valid for P1, but there is some observational evidence that the most metal-rich ([Fe/H]>-1.2) stars in P2 may have formed later, subsequent to some loss of gas from the system. This can be seen in the ChM (Fig. \ref{fig:chm}), where there is a lack of stars connecting the two overdensities that trace our P2 stars. Such a discontinuity is also seen in M54 \citep{milone_atlas_chm}, a system known to have undergone a bursty history of star formation. Fig. \ref{fig:chm} shows that there is a significant discontinuity between the most metal-rich P2 stars and the main body of the P2 stars on the ChM. This may in fact reflect such a pause in the SFR, during which time some of the gas was removed. 

\subsubsection{Models without MDF constraints: a new mass budget problem}\label{sec:unconstrained}

In this Section we examine the performance of models optimised without the imposition of an MDF constraint.  {The resulting histories of gas infall and star formation, as well as the evolution of [Fe/H] with time, are shown as dotted lines in Fig.~\ref{fig:sfrs_p1_p2}.
Comparisons with data are } displayed in the bottom panels of Figs.~\ref{fig:alpha_fe_p1} and \ref{fig:alpha_fe_p2}.  
For both P1 and P2 populations, the match to the run of [Mg/Fe] vs [Fe/H] is only marginally improved---a little more so in the case of P1, whose metallicity now extends beyond [Fe/H]$\simeq$--1.2.
Unsurprisingly, the biggest changes take place in the MDF predictions, which in both cases get shifted to greater power towards the high metallicity end. 
The variation is more extreme in the case of P1, which presented a fairly good match in the ``constrained'' case, and now displays a sizeable wing towards [Fe/H]$\simgreater$--1.5.  
In the case of P2, most of the power is now located at [Fe/H]$\simgreater$--1.2, in sharp constrast with the  observed MDF.

These results suggest the presence of an inconsistency between the apparent evolution of both populations in the Mg-Fe plane and their MDFs.  
They suggest that, in order to produce the strong increase in [Mg/Fe] observed towards the metal-rich half of both populations, the system must undergo a strong burst of star formation, thus producing an over-abundance of metal-rich stars which are not observed.
This new type of {\it mass budget problem} manifest in our models can be explained in two possible ways.  In one scenario, it can be argued that these metal-rich stars were actually formed, but due to their being somehow less bound to the system they were tidal-stripped through interaction with the Milky Way host halo.
Alternatively, gas stripping could be responsible for the mismatch between the predicted and observed MDFs.  The prescription for outflows due to stellar feedback (eq.~\ref{eq:outflows}) are likely too simplistic to represent a situation where gas may have been tidally stripped through interaction with the Milky Way halo. {In addition, the model does not allow for temporary quenching of star formation.}
Indeed, continued interaction between an infalling satellite galaxy and the host halo can contribute to {the gradual removal of the star forming gas}, as well as star formation bursts.  Such events may single-handedly explain the {discrepancies observed between the model and observed MDFs we showed in Figs.~\ref{fig:alpha_fe_p1} and \ref{fig:alpha_fe_p2}.}

{The model histories of star formation displayed in Fig.~\ref{fig:sfrs_p1_p2} predict an age spread of $\simeq 4$~Gyr for both populations, which is in good qualitative agreement with the recent determinations by \cite{clontz2024}}. 
In both cases, the burst of star formation is predicted to have occurred in the latest stages of the chemical evolution of both populations, being associated with the formation of their most metal-rich stars. Interestingly, it is in this regime that important discrepancies between observed and predicted MDFs are found. We further elaborate on this result in Section~\ref{sec:discussion}.

\bigskip

{
\subsection{Summary of results from GCE modelling}\label{sec:model_summary}
}

{ Employing the {\tt VICE} GCE modelling code, we have derived best-fitting GCE models  to the P1 and P2 populations in $\omega$ Cen, under the hypothesis that these populations evolved in chemical detachment. This assumption is well-motivated by an inspection of the chemical abundance patterns evident in Figure~\ref{fig:6panel_abundances}.
By optimising parameters dictating the evolution of gas infall, outflows, and star formation efficiency, we find that the models predict that both populations underwent a burst of star formation, preceded by a few Gyr of low star formation rate. 
The GCE models are a good match to the distribution of both P1 and P2 stars on the Mg-Fe plane, but fail to reproduce their MDFs. 
The mismatch is of course exacerbated when MDFs are not adopted as constraints in the optimisation.  In both cases the best-fitting models predict an excess of metal-rich populations formed during the bursts of star formation}. 

In the following section, we first compare the data and our interpretations to other { studies of the chemical composition of $\omega$ Cen stars.
Following that, we} perform a chemical comparison between the P1 stars of $\omega$ Cen and the field stars of accreted dwarf galaxies in the Milky Way's stellar halo in an attempt to constrain { the nature of} $\omega$ Cen's parent population. 
Finally, we tie in our chemical tagging of the P1, P2 and IM stars in the VAC with the larger observational state of play concerning $\omega$ Cen. 

\bigskip
\bigskip

\section{The Multiple Populations of \texorpdfstring{$\omega$}~~Cen in the context of chemical evolution models}\label{sec:discussion}



In $\S$\ref{sec:chemistry} and \ref{sec:gce_modelling}, we  {explored new insights into the mix of stellar populations hosted by the $\omega$~Cen stellar system}, afforded by a combination of precise chemical abundances from the APOGEE VAC, and the oMEGACat photometry.  {We showed that APOGEE chemistry splits the stars of $\omega$~Cen into three stellar populations, which are mapped neatly onto separate} sequences on the chromosome map. 
Furthermore, we presented exploratory galaxy chemical evolution models that attempt to explain these abundance patterns in terms of episodes of star formation occurring in star-forming gas reservoirs with markedly different compositions. 

In this section, we discuss the shortcomings of the GCE models described in $\S$\ref{sec:gce_modelling}, particularly in relation to the MDF of P2 and the implications of the model's failure to reproduce it. Following that, we speculate on the origin of the intermediate and P2 populations, which are both characterised by the anomalous abundance patterns associated with the Galactic GCs. 
Finally, under the assumption that the P1 sample comprises former field stars of the progenitor of $\omega$ Cen, we perform chemical comparisons between these stars and other stars belonging to substructure identified in the literature, particularly ones where there has been a speculative link to $\omega$ Cen's progenitor.

\subsection{On the role of gas mass loss in shaping the MDF of P2}

In $\S$\ref{sec:gce_modelling}, we fitted GCE models {to Mg and Fe abundances} of the P1 and P2 samples constructed in $\S$\ref{sec:data}. 
The underlying assumptions of these models are that after a period of initially inefficient star formation, {a starburst takes place} without the inflow of additional gas, due to a sudden increase in the star formation efficiency. 
Such a starburst (dubbed an `efficiency-driven starburst' \citealp[e.g.,][]{nidever_lmc, vice1} is markedly different from one driven by accretion, as it simply marks an enhancement in the consumption of the available gas, enhancing the rate of SN II enrichment relative to that of SN Ia from antecedent star formation. Such a burst could be driven by a dynamical disturbance to the existing gas supply.

What Fig. \ref{fig:alpha_fe_p2} shows is that at the onset of the starburst, there is too much gas available in the reservoir and thus the number of stars at [Fe/H]$\gtrsim-1.2$ is vastly over-predicted by the model. 
Fig. \ref{fig:chm} indicates that the sequence on the ChM corresponding to our P2 sample is characterised by a dearth of stars between $\Delta_{\rm F275W,~F814W}\simeq0.7$ and $\Delta_{\rm F275W,~F814W}\simeq1.0$. 
The stars comprising the overdensity at $\Delta_{\rm F275W,~F814W}\simeq1.25$ are located on the locus on the Mg-Fe plane that must have formed during the starburst, characterised by constant [Si/Fe], enhancement in [Mg/Fe], and enhanced [N/Fe] as a function of [Fe/H] (see stars with [Fe/H]$\gtrsim-1.2$ in Fig. \ref{fig:6panel_abundances}). 
This `gap' on the ChM is also seen in M54 \citep{milone_atlas_chm}, a system understood to have experienced a bursty history of star formation and boasting metal-rich populations like $\omega$ Cen \citep{bellazzini_08_m54}, combined with the gradual stripping of its gas during its interaction with the Milky Way. 
Thus, we conclude that there may have been a temporary period of low-intensity or quenched star formation, during which a significant fraction of the gas reservoir was removed from $\omega$ Cen. A subsequent burst of star formation, ocurring in the now significantly-depleted gas reservoir, formed the metal-rich stars of P2.  
{Since our GCE models adopt a simplistic feedback-motivated outflow, gas removal is underestimated, and as a result they badly overestimate the number of metal-rich stars in the P2 population. }

\subsection{The origin of the IM population}
\label{sec:origin_im}

In $\S$\ref{sec:chemistry} we compared the abundance patterns of the intermediate population to the {so called ``second generation''} populations of Galactic GCs with similar metallicity. 
We found that not only are their metallicity spreads consistent with those of the IM population, they also show the same abundance anticorrelations expected of chemically anomalous populations in those GCs. Fig. 
\ref{fig:6panel_abundances_mg} also indicates that the P2 and IM populations in $\omega$ Cen exhibit significantly different abundance patterns, with P2 being more enhanced in Si and Al.


We speculate that this population is the result of GCs spiralling into the centre of $\omega$ Cen at early times. 
$\omega$ Cen has long been speculated to be a nuclear star cluster, the nucleated remnant of a satellite that merged with the Galaxy. 
These systems are hypothesised to grow by the spiralling in of the host galaxy's field clusters by dynamical friction, in-situ star formation, or { most likely} a combination of both. 
Assuming a typical ratio between NSC and host galaxy mass, $\omega$~Cen was likely hosted by a galaxy with $M_\star\simeq10^9~{\rm M_\odot}$ which is in the mass regime where both processes contribute to NSC growth \citep{fahrion_nscs_19}. 
Thus, given its apparent chemical disconnect with the P1 and P2 populations and the fact that it exhibits the standard Mg-Al anti-correlation, it is reasonable to assume that the IM population originates from the inspiralling of metal-poor GCs towards the centre of $\omega$~Cen's host galaxy.


{
\subsection{On the origin of the P1 and P2 populations}
}

Having a working hypothesis for how the IM population has happened upon $\omega$~Cen, we now turn to an interpretation of our results for the P1 and P2 populations. A critical aspect of our approach is that we choose to model the chemical evolution of the two populations separately, without attempting to establish a chemical evolution link between them.  
By proceeding in that way we renounce any ambition to devise a fully consistent model for the chemodynamic evolution of the $\omega$~Cen stellar system.  Indeed, formulating a chemical link between populations such as P1 and P2 is a proposition that has eluded the community for well over a decade \citep[see the discussion by][]{bastian_gc_review}, and is beyond the scope of this work. 
Our focus instead is on understanding what type of star formation and chemical evolution histories can produce such a unique distribution of chemical properties as observed in $\omega$~Cen.  In doing so we hope to gain new insights into the history of this peculiar system.

We start by looking at the evidence for the occurrence of star forming bursts in these systems.  As discussed in Sections~\ref{sec:chemistry} and \ref{sec:gce_modelling}, a steep relation between abundance ratios such as [Mg/Fe] and [Al/Fe] and [Fe/H] is a telltale sign of a burst of star formation, as it implies a preponderance of enrichment by massive stars \citep[e.g.,][]{hasselquist_lgds,fernandes2023}. 
Indeed, optimisation of GCE model parameters using the {\tt VICE} code results in SFHs characterised by low level star formation followed by a strong burst for both P1 and P2 (Fig.~\ref{fig:sfrs_p1_p2}). It is important to note that the starting time for both models is completely arbitrary, so that age differences implied by the SFHs displayed in Fig.~\ref{fig:sfrs_p1_p2}) are difficult to interpret.

{Bursts of star formation can be triggered by interactions between infalling satellites and their hosts \cite[e.g.,][]{bekki_wcen,emsellem08_nsc,pearson2019}.  
In addition, such episodes of star formation in infalling satellites are accompanied by gas stripping through various processes, such as tidal forces, dynamical friction, and stellar feedback \citep{bassino94, pfeffer13}.
In this context, it may be possible to understand the MDF mismatch displayed in Figs.~\ref{fig:alpha_fe_p1} and \ref{fig:alpha_fe_p2} as a by-product of the merger process just as much as the bursts of star formation themselves.  
In other words, the deficit in metal-rich stars (particularly important in the case of P2), may be the result of the loss of metal-rich gas incurred during the accretion of the $\omega$~Cen host into the halo of the Milky Way. 
This gas mass deficit would then account for the reduced impact of the star formation burst on the final stellar mass budget of $\omega$~Cen, explaining the MDF mismatches.
}

{
\subsection{Putting the pieces together: a hypothetical scenario for the genesis of $\omega$~Cen}\label{sec:origin_p1p2}
}

Before proceeding, it is suitable that we take stock of where we are with the different pieces of the puzzle.  According to our k-means analysis (Section~\ref{sec:chemistry}),  $\omega$~Cen hosts three stellar populations, characterised by distinct chemistry.  One population (IM) is likely the result of the inspiralling of one or more metal-poor GCs into the centre of the $\omega$~Cen host galaxy.
The remaining populations are P1, which is characterised by abundance ratios that are akin to those of halo field stars at same metallicity, whereas P2 displays extreme second-generation GC chemistry at a broad range of metallicities.  
The two populations seem to have undergone separate histories of star formation and chemical enrichment, both characterised by a period of ``simmering'' star formation, followed by a starburst.  

Since the P1 burst is triggered at a time when the gas has substantially lower metallicity than P2 ([Fe/H]$\approx$--1.7 as opposed to $\approx$--1.3), it is reasonable to suppose that the P1 burst took place at an earlier time.  This notion is further supported by the fact that the models predict a longer period of ``simmering'' star formation for P2 than for P1 (Fig.~\ref{fig:sfrs_p1_p2}).  Thus the evidence favours the P1 burst having occurred first. 


A possible scenario would thus start with P1 as the direct chemical descendant of the primordial stellar population residing in the centre of the $\omega$~Cen host galaxy.  Steady conversion of gas into stars at a low rate proceeded until the falling into the Milky Way halo triggered a burst of star formation.  
Interaction with the Milky Way could then have led to a quenching of the star formation rate, due to gas loss associated with tidal stripping, harassment, and/or feedback. 

The P1 hypothesis being accepted, one is then left with the difficult question regarding the origin of the gas from which the P2 population was formed. As mentioned above, this is a fundamental unsolved problem in the present understanding of GC formation \citep[see, e.g.,][]{renzini2015,bastian_gc_review}.
It is beyond the scope of this paper to attempt a solution, so we simply take the existence of multiple populations in GCs for granted. What follows are mere speculations based on the information at hand. 

We have assumed that the P1 and P2 populations evolved in chemical detachment. At first glance this assumption may seem unreasonable in view of the fact that these two populations are tightly co-located today in a dense environment. 
The obvious competing scenario would be one according to which the system underwent accretion of gas with the chemical composition needed to, upon mixing with the existing {\it in situ} gas, dilute its chemical composition so as to next form stars with the abundance ratios observed in the metal-poor end of the P2 population.
Looking at Figure~\ref{fig:6panel_abundances_mg}, that would require, for instance, a decrease of $\simeq$1.6~dex in [Mg/H] and 3 dex in [Al/H]. In short, the chemistry of the early P2 populations is so exceptional that for it to result from mixing with pre-existing evolved P1 gas would call for infall gas abundances that may be unreasonably extreme. While worth mentioning it, we deem this scenario unlikely.

A possible source for the gas that formed P2 is the inspiralling GC(s) that gave origin to the IM population. Since the IM population contains stars with second-generation chemical compositions, it is conceivable that the inspiralling of their host GC(s) brought gas whose abundances were characteristic of that extreme abundance pattern. Such stars are also found in even larger amounts in the field \citep[e.g.,][]{schiavon2017a,trincado2019,kisku2021,horta2021,phillips2022,belokurov2023}. A similar scenario has been proposed by \cite{alvarez_garay_wcen} as a mechanism to build up $\omega$ Centauri's MPs.

While at present we lack a clear definition of what process is responsible for this phenomenon, there is no question about its ability to generate enough gas to form $\approx 10^6~{\rm M_\odot}$ in 2G stars within the most massive GCs. 
These inspiralling GC(s) would  likely be forming the early P2 stars at a low level of star formation, but then infall of such a large amount of dense gas into the core of $\omega$~Cen's host galaxy may trigger a second burst of star formation, responsible for the production of the metal-rich P2 stars observed in $\omega$~Cen today. 
Stripping of that gas partly during the inspiralling into the host galaxy, and partly due to feedback and harassment by the Milky Way may be responsible for the ``metal-rich mass budget problem'' laid bare by the MDF comparisons of Fig.~\ref{fig:alpha_fe_p2}.




\bigskip

\section{Can we identify the remains of \texorpdfstring{$\omega$}~~Cen's host galaxy on the basis of chemistry?}

As we discussed in $\S$\ref{sec:intro}, there may be a genetic link between $\omega$ Cen and accreted populations in the Milky Way's stellar halo. The two most prominent candidates suggested to date for $\omega$ Cen's former host system are the Sausage/Gaia Enceladus \citep{belokurov_sausage, helmi_ges} and the Sequoia \citep{Myeong_sequoia}.

\cite{horta23_halosubs} utilised a $\chi^2$ method on the basis of APOGEE data to compare the chemical abundance patterns of substructures in the Milky Way stellar halo to {\it in situ} stars at the same metallicity. 
Notable inclusions in the analysis were Heracles \citep{horta21_heracles}, the Sausage/Gaia-Enceladus \citep{belokurov_sausage,helmi_ges,haywood2018,mackereth2019}, the Sagittarius dSph \citep{ibata_m54}, the Helmi stream \citep{helmi99_streams}, and Sequoia \citep{barba19_sequoia, Myeong_sequoia}.

We perform the same exercise, comparing the abundances of the Sausage/Gaia-Enceladus, Sequoia, Heracles, and Aurora \citep{belokurov22_aurora, myeong22_aurora} to the P1 stars in the VAC. 
Aurora is purportedly the {\it in situ} relic of the Milky Way prior to the onset of the formation of the disk, characterised by hot kinematics, an isotropic velocity ellipsoid, and slight rotation. 
The comparison is based on the assumption that P1 stars {constitute the original} field stars of $\omega$~Cen's progenitor, since no halo substructure identified to date is dominated by stars with chemistry similar to P2 or IM. 

To briefly summarise the method presented in \cite{horta23_halosubs}, for each population considered we make corrections to the abundances to account for systematic abundance variations with surface gravity ($\log{\rm g}$), which can be caused by a combination of stellar evolution or systematic effects as a function of stellar parameters \citep[see][for a thorough discussion]{weinberg21_2proc}. 
We make these corrections, restricting our sample to stars with $1<\log{\rm g}<2$, and fit second order polynomials to the relationship between $\log{\rm g}$ and [X/H] for every species entering the comparison (X$\in[{\rm O, Mg, Si, S, Ca, Ti, C, N, Al, K, Mn, Ni, Ce}]$. Corrections are made by subtracting the difference between the polynomial fit to the data and the observed abundance. 

{Next we determined, for each substructure, the dispersion of the abundance ratios [X/Fe] at two reference metallicities, [Fe/H]$_{\rm comp}=-1.7$ and $-1.2$.  
That was achieved through a boostrapping resampling method, generating 1000 realisations of the X-Fe planes of every substructure considered, for every species X, selecting stars within a $\pm0.1$~dex window around those two reference [Fe/H] values.}
This yields 1000 median [X/Fe] values for the 13 elements adopted for the comparison, from which we compute the mean and standard deviations of [X/Fe], using them to compare the chemical abundances between the two populations.

{Using the mean and dispersion values computed for every [X/Fe] at the two ${\rm [Fe/H]_{\rm comp}}$ values, we assess the chemical similarity between P1 and halo substructures by using a $\chi^2$ statistic given by Eq.~1 from \cite{horta23_halosubs}:}

\begin{equation}
    \chi^2 = \sum_{i} \frac{({\rm [X/Fe]_{i, {\rm sub}}} - {\rm [X/Fe]}_{i,{\rm P1}})^2}{(\sigma^2_{{\rm [X/Fe]}_{i, {\rm sub}}} + \sigma^2_{{\rm [X/Fe]}_{i, {\rm P1}}})},
    \label{equation_chi2}
\end{equation}
where ${\rm [X/Fe]_{i, {\rm sub}}}$ and ${\rm [X/Fe]}_{i,{\rm P1}}$ are the abundances of a given halo substructure and the P1 stars, respectively. $\sigma^2_{{\rm [X/Fe]}_{i, {\rm sub}}}$ and $ \sigma^2_{{\rm [X/Fe]}_{i, {\rm P1}}}$ are the corresponding uncertainties for those abundance measurements.
We then compute the $p$-value for the $\chi^2$ statistic for 12 degrees of freedom using {\texttt{scipy}}'s {\texttt{scipy.stats.chi2.cdf}} routine, {where a value of $p_{\chi^2}<0.05$ indicates that the abundances of P1 and halo substructures are not drawn from the same population.} 
Finally, we also compute the sum of the differences, $\sum\Delta_{\rm [X/Fe]}$, given by the numerator of Eq. \ref{equation_chi2}.

{The samples and data for each substructure came from \cite{horta23_halosubs}}  The sample of {\it in-situ} stars (represented by black points and tracks) was chosen to mimic the same selection in \cite{conroy_h3_disk} (who adopted a left-handed coordinate frame, hence we select stars with $L_z~>~500~{\rm kms^{-1}kpc^{-1}}$ and $e<0.8$).

\begin{figure*}
    \centering
    \includegraphics[width = \textwidth]{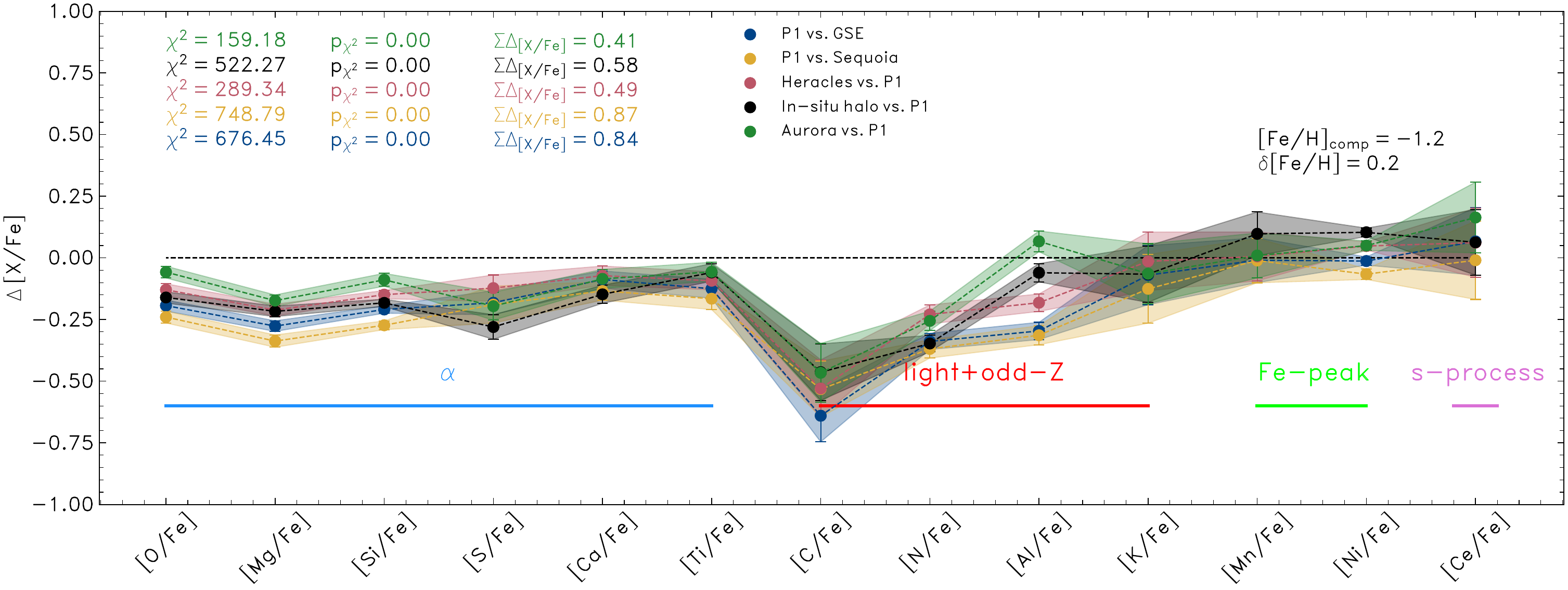}
     \caption{$\Delta$[X/Fe] differences (Sub - P1) between the resulting mean values obtained using the method presented in $\S$5 of \protect\cite{horta23_halosubs} in 13 different chemical abundance planes at [Fe/H]=-1.2$\pm\sigma_{\rm [Fe/H]_{P1}}$. Here we compare the stars comprising P1 to the {\it i)} Aurora (green) {\it ii)} Sausage/Gaia Enceladus (navy), {\it iii)} Sequoia, {\it iv)} Heracles, and {\it v)} a sample of {\it in-situ} halo stars.}
    \label{fig_chemical_comp_2}
\end{figure*}

Fig. \ref{fig_chemical_comp_2} shows the results for $\rm [Fe/H]_{\rm comp}=-1.2\pm0.2$.  In this metallicity none of the substructures exhibits the same abundance pattern as P1. 
This is not surprising. For example, at this [Fe/H]$_{\rm comp}$ P1 exhibits $\upalpha$-abundances $\approx$~0.3 dex higher than those in these substructures at the same metallicity ($\rm [Mg/Fe]\approx0.5$). 



{When running the same statistical test at [Fe/H]$_{\rm comp}=-1.7$ we find smaller discrepancies between all substructures and the P1 population (we skipped Aurora, because the sample does not reach low enough [Fe/H] for a meaningful comparison).  In fact, for Heracles we find formal similarity with P1 ($p_{\chi^2}=0.14$).  
One could reasonably argue that these abundance differences are the result of chemical composition gradients within the host galaxy. 
However, the one case for which chemistry of the body, outskirts, and NSC of a satellite galaxy are available, Sgr dSph, an [$\upalpha$]/Fe] gradient is not present \citep{hayes2020}.}


We thus conclude, on the basis of this analysis, that none of the halo substructures contained in the APOGEE DR17 catalogue for which an association with $\omega$~Cen has been claimed \citep[e.g.][]{limberg_sausage} has chemical compositions that are consistent with such an association.

{\section{Open questions} \label{sec:openq}

The highly speculative scenario presented in $\S$\ref{sec:origin_im} and $\S$\ref{sec:origin_p1p2} accounts for some of the broad properties of the $\omega$~Cen stellar system, but leaves a number of questions unanswered.  It is critical that they are stated clearly, and we enumerate them below.

\begin{enumerate}
    \item {\it How was the P2 gas originally enriched?} Our scenario for the origin of the P2 populations suffers from a fundamental shortcoming. In order to fit the chemical evolution of that population we had to {\it assume} that it started from gas that was originally enriched to a somewhat extreme second-generation GC abundance pattern. That assumption, while justified by the observations, is devoid of a theoretical foundation.  Although we know that all GC second-generation stars must have formed from gas characterised by such extreme chemical compositions, there currently is no model capable of producing it while matching all the properties of the multiple populations in GCs \citep[e.g.,][]{bastian_gc_review}. Without an answer to this basic question, a definitive picture for the origin of $\omega$~Cen's populations---and that of that perplexing stellar system itself---will still remain elusive. 

    \item {\it How much mass did $\omega$~Cen lose?} Consider the {\it metal-rich mass budget problem} discussed in Section~\ref{sec:unconstrained}. Our scenario explains away the metal-rich MDF mismatch (Figs.~\ref{fig:alpha_fe_p1} and \ref{fig:alpha_fe_p2}) as being caused predominantly by loss of gas due to tidal stripping and harassment.  Nevertheless, there is strong evidence that $\omega$~Cen lost substantial mass in the form of stars \citep[e.g., Anguiano et al., 2025, submitted,][]{pagnini2025,galah_fimbulthul,ibata_fimbulthul_wcen}. 
    A reliable estimate of the amount of stellar mass lost by $\omega$~Cen over the past many Gyr will have to await the chemical tagging of a statistically robust halo field sample.  On the other hand, we have no means of ascertaining the total gas mass lost.  However, if the model predictions displayed in Figs.~\ref{fig:alpha_fe_p1} and \ref{fig:alpha_fe_p2} are accurate, one would reasonably conclude that it lost most of its mass to the field, predominantly in the form of P2 stars and/or gas.

    \item {\it Are the extreme abundance ratios of the P2 population a feature of Nuclear Star Clusters?}  We hypothesize that the P2 population is the result of the conversion of 2G gas present in the GC(s) that spiralled into the centre of $\omega$~Cen's host galaxy. Figs.~\ref{fig:6panel_abundances} and \ref{fig:6panel_abundances_mg} show that P2 star formation starts from gas with characteristically low [Mg/Fe] and very high [Al/Fe].  
    This is a regime found in very few Galactic GCs \citep[see Fig.~8 of][]{vac_paper}. It may be reasonable to assume that such extreme abundance patterns are the result of star formation in GCs that are under the effect of a strong interaction with the galaxy host they are spiralling into.  
    Perhaps GCs that are not nuclear clusters never manage to enrich the intracluster gas to such extreme abundance levels.  If that is the case, one could reasonably hypothesize that the abundance patterns of 2G stars in NCs (our P2 population) constitute an upper limit on the abundance ratios attained by 2G stars in normal GCs.  
    If that is correct, our results may place important constraints on the source of Al-enrichment/Mg-depletion in GCs.  Moreover, if the run of Al with metallicity seen in  Figs.~\ref{fig:6panel_abundances} and \ref{fig:6panel_abundances_mg} can be understood on theoretical grounds, one would be able to explain why multiple populations in metal-rich GCs do not attain a wide range of Al abundances \citep[e.g.,][]{schiavon2017b,nataf2019,vac_paper}. 
    \item{\it What do the differences in Si/Mg between P2 and P1/IM mean?} Fig.~\ref{fig:6panel_abundances_mg} shows that the abundance ratio [Si/Mg] is much higher in P2 than in either P1 or IM populations.  As pointed out by \cite{carlin2018}, this ratio is sensitive to the initial mass function (IMF) of the system.  
    That is because Si is produced in explosive nucleosynthesis by SN~IIe, whereas Mg is produced during hydrostatic nuclear burning in massive stars. As a result, yields of these two elements are a function of stellar mass.
    In particular, Mg is only produced in stars with high enough masses that hydrostatic burning of C and Ne is ignited. Therefore, that P2 has a much higher [Si/Mg] ratio than P1 and IM may suggest, all other relevant quantities being the same, that the IMF of P2 is top light, compared with that of P1 and IM.  
    This observation may be related to the fact that star formation in P2 took place in a considerably different environment than P1.  Finally , it is also possible that SN~Ia contribution to the enrichment of Si \citep{kobayashi2020} may play a role in this conundrum.


\end{enumerate}

}

\section{Summary}

We have selected and examined the stellar distributions on canonical chemical planes of the multiple populations hosted by $\omega$ Centauri using the APOGEE Value-added Catalogue of Galactic globular clusters \citep{vac_paper}. In doing so, we have placed constraints on the assembly history of this complex stellar system. Furthermore, in our cross-match with oMEGACat and construction of the ChM we are able to tie our interpretations into the wider observational state of play. Our main results can be summarised as follows.

\begin{enumerate}
    \item { Application of standard k-means substructure finding to the abundances of Fe, Mg, Si, Al, and Mn, leads to the identification of three distinct populations in $\omega$~Cen. 
    The so-called P1 and P2 populations display a broad distribution of metallicities and strong correlations between abundance ratios of Mg, Si, Al, and N and metallicity.  Such correlations are strong signatures of chemical evolution of the gas forming these two separate populations.  The so-called IM population has a narrower range of metallicities, is metal-poor, and displays the abundance anti-correlations commonly present in GCs.}  

{
    \item Matching the APOGEE/VAC sample to photometry from the oMEGACat survey, we mapped the loci of the three populations on the so-called chromosome map (ChM) for $\omega$~Cen.  We find that the P1, P2, and IM populations span the entirety of the area covered by $\omega$~Cen stars in the ChM.  
    The P1 and P2 sequences connect multiple density peaks within the ChM, which consist of stars with different [Fe/H], but similar abundances of light elements.
    We propose that these peaks, which are typically associated with distinct stellar populations in the literature, are instead connected by a history of star formation and chemical evolution.  
    We thus conclude that the chemical complexity of $\omega$~Cen stars can be accounted for by the chemistry of the IM, P1, and P2 populations, and in particular the chemical evolution of the latter two.

}

    \item The chemical compositions of P1 stars are similar (but not identical) to those of dwarf galaxies and the stellar halo at the same [Fe/H]. Starting at ${\rm [Fe/H]\approx-1.8}$, [Al/Fe] and [$\upalpha$/Fe]  show an increasing trend with respect to [Fe/H], with the latter showing no decline characteristic of the $\upalpha$ knee.  P2's chemical compositions are characteristic of the most extreme populations seen in Galactic globular clusters. It has significant Al-enhancement (reaching ${\rm [Al/Fe]\approx +1.2}$); Si-enhancement (as high as ${\rm [Si/Fe]\approx +0.4}$), and significant Mg-depletion (as low as ${\rm [Mg/Fe]\approx -0.4}$). Its abundance patterns are characterised by increasing [Mg/Fe], declining [Al/Mg], and constant [Si/Fe] as a function of [Fe/H].  The IM population has a much smaller spread in [Fe/H], and it displays the standard Mg-Al anti-correlation typical of metal-poor Galactic globular clusters.  

{
    \item By assuming that the P1 and P2 populations evolve in chemical detachment, we run models of galactic chemical evolution using the {\tt VICE} package to match the behaviour of these populations in the Mg-Fe chemical plane. The best fitting models for both populations consist of a history of star formation characterised by a starburst preceded by a few to several years of low level star formation.  The models are a good match to the data on the Mg-Fe plane.  We hypothesise that the density peaks identified in the ChM along the P1 and P2 sequences are associated with bursts of star formation that are not represented in our model star formation history because the APOGEE MDF is too sparse to resolve them.
}

{
    \item Knowing that $\omega$~Cen has lost a large amount of stellar mass in its past, we run a {\tt VICE} optimisation that ignores the MDFs of both P1 and P2. The resulting predicted MDFs contain far more power in the metal-rich end than observed. This ``metal-rich mass budget problem'' implies selective loss of stars and/or gas on the high metallicity end, predominantly by the P2 population.
}

{
    \item We propose a strawman scenario according to which  the P1 population was formed first, as a result of chemical evolution from primordial populations in the centre of $\omega$~Cen's host galaxy.  The IM population is the result of the spiralling in metal-poor globular clusters towards the centre of the host galaxy of $\omega$~Cen.  P2 may form from gas enriched to extreme 2G chemical composition levels within the GC(s) that became the IM population.  We speculate that, through this process, extreme 2G abundance patterns such as those seen in P2 are a feature exclusive of nuclear star clusters.
}

{
    \item The ratio of hydrostatic to explosive $\alpha$-elements in P2 is much lower than that in P1 and IM. This may be due to P2 having had a top-light IMF.
}

{
    \item Finally, we run a robust comparison of the detailed chemical composition of the P1 population with those of halo field substructures Sausage/Gaia-Enceladus, Sequoia, Heracles, and  Aurora. The data suggest no chemical association between $\omega$~Cen and any of those substructures.

}

\end{enumerate}

\section*{Acknowledgements}

We thank David Weinberg, Sten Hasselquist, Maurizio Salaris, Nate Bastian, Nadine Neumayer, Anil Seth, and Selina Nitschai for helpful discussions during preparation of this paper.

Analyses and plots presented in this article used {\textsc{IPYTHON}} and packages in the {\textsc{SCIPY}} ecosystem (\citealp{jones_scipy, hunter_matplotlib,perez_ipython,seabold_statsmodels,vanderwalt_numpy}).

SKA gratefully acknowledges funding from UKRI through a Future Leaders Fellowship (grants MR/T022868/1,
MR/Y034147/1).

SS acknowledges funding from the European Union under the grant ERC-2022-AdG, {\em "StarDance: the non-canonical evolution of stars in clusters"}, Grant Agreement 101093572, PI: E. Pancino.
\section*{Data Availability}

All APOGEE DR17 data upon which this study are based are publicly available and can be found at \url{https://www.sdss4.org/dr17/}.



\bibliographystyle{mnras}
\bibliography{working_draft} 

\bsp	
\label{lastpage}
\end{document}